\documentclass[11pt,preprint]{elsarticle}
\usepackage{float}
\usepackage{apalike}
\usepackage{amsmath,amssymb,amsfonts}
\usepackage{multirow}
\usepackage{booktabs}
\usepackage{graphicx}
\usepackage{threeparttable}
\usepackage{lscape}
\usepackage{adjustbox}
\usepackage{siunitx}
\usepackage{bbm}

\usepackage[ruled,norelsize]{algorithm2e}
\SetAlFnt{\small}
\SetAlCapNameFnt{\small}

\newcommand{\argmin}{\mathop{\rm arg~min}\limits}

\journal{arXiv}

\begin{document}
\begin{frontmatter}

		
		
		
		
		
		
		

\title{EMG Pattern Recognition via Bayesian Inference with Scale Mixture-Based Stochastic Generative Models}

\author[label1]{Akira Furui\corref{cor1}}
\ead{akirafurui@hiroshima-u.ac.jp}

\author[label2]{Takuya Igaue}

\author[label1]{Toshio Tsuji\corref{cor1}}

\nonumnote{This paper is an accepted version for publication in Expert Systems with Applications.}
\cortext[cor1]{Corresponding authors.}
\address[label1]{Graduate School of Advanced Science and Engineering, Hiroshima University, Higashi-hiroshima 739-8527, Japan}
\address[label2]{Graduate School of Engineering, The University of Tokyo, Bunkyo-ku 113-8656, Japan}

\begin{abstract}
Electromyogram (EMG) has been utilized to interface signals for prosthetic hands and information devices owing to its ability to reflect human motion intentions.
Although various EMG classification methods have been introduced into EMG-based control systems, they do not fully consider the stochastic characteristics of EMG signals.
This paper proposes an EMG pattern classification method incorporating a scale mixture-based generative model.
A scale mixture model is a stochastic EMG model in which the EMG variance is considered as a random variable, enabling the representation of uncertainty in the variance.
This model is extended in this study and utilized for EMG pattern classification.
The proposed method is trained by variational Bayesian learning, thereby allowing the automatic determination of the model complexity.
Furthermore, to optimize the hyperparameters of the proposed method with a partial discriminative approach, a mutual information-based determination method is introduced.
Simulation and EMG analysis experiments demonstrated the relationship between the hyperparameters and classification accuracy of the proposed method as well as the validity of the proposed method.
The comparison using public EMG datasets revealed that the proposed method outperformed the various conventional classifiers.
These results indicated the validity of the proposed method and its applicability to EMG-based control systems.
In EMG pattern recognition, a classifier based on a generative model that reflects the stochastic characteristics of EMG signals can outperform the conventional general-purpose classifier.
\end{abstract}

\begin{keyword}
Electromyogram (EMG) \sep pattern recognition \sep motion classification \sep scale mixture model \sep Bayesian inference.
\end{keyword}

\end{frontmatter}

\section{Introduction}
Biological signals reflecting the internal state of humans are useful in realizing natural and intuitive  human-machine interfaces.
In particular, electromyogram (EMG) signals produced by muscle contractions have been widely studied because their patterns can be voluntarily controlled by users.
EMG signals are recorded as the summations of individual action potentials generated by motor units.
Motion determination based on EMG pattern recognition is one of the most representative applications, and many attempts have been made to develop EMG-based control interfaces, such as prosthetic hands~\citep{Farina2017-rr,Furui2019-bz} and information devices~\citep{Lu2019-le}.

To enable such applications, it is important to extract accurately the motion intention of the operator from the measured EMG signals.
This objective is achieved by estimating the correspondence between the motion and EMG patterns of the operator based on machine learning.
Thus, our goal is to enable better prediction of the posterior distribution $p(c|\mathbf{x})$ with respect to the class label $c$ of the motion and the EMG pattern $\mathbf{x}$.
Classifiers that have been used for EMG pattern recognition can be categorized into two types: discriminative classifiers and generative classifiers.

Discriminative classifiers learn the decision boundaries themselves from the data by directly modeling the output, $p(c|\mathbf{x})$.
This approach is generally highly versatile and flexible because it does not require the explicit assumption of the distribution that the input data follow.
Many discriminative models have been adopted for EMG classification; typical examples are the multilayer perceptron (MLP)~\citep{Hudgins1993-sa,Bird2020-yc,Dellacasa_Bellingegni2017-ke}, support vector machine (SVM)~\citep{Oskoei2008-wz,Liu2015-re,Dellacasa_Bellingegni2017-ke,Li2020-va}, and $k$-nearest neighbors ($k$-NN)~\citep{Kim2011-gp,Nishad2019-nv}.
Although these classifiers have high representation capabilities, it is difficult to introduce task-specific knowledge regarding the data generation process into these classifiers and they require appropriate constraints on their structures to prevent overfitting, which often involves heuristic settings.

Alternatively, generative classifiers model the joint distribution $p(\mathbf{x}, c)$ and subsequently use this joint distribution to calculate the posterior $p(c|\mathbf{x})$.
This approach typically involves stronger assumptions regarding the data generation process because it requires the distribution of the inputs, as well as that of the outputs, to be designed.
It has been shown that if this assumption is appropriate, i.e., if the mismatch between the data distribution and model is small, then generative classifiers exhibit better performance than that of discriminative classifiers~\citep{Ng2002-sv,Bishop2007-pd}.
Several attempts have been made to model the stochastic properties of EMG signals, which are generally assumed to follow a Gaussian distribution~\citep{Parker1977-hl,Hogan1980-pw,Abbink1998-oh}.
Following this assumption, Gaussian distribution-based classifiers, such as the Gaussian mixture model (GMM) and linear discriminant analysis (LDA), have been commonly introduced into EMG-based control systems.
Chan and Englehart focused on this generative approach and demonstrated the potential of the GMM, a linear combination of Gaussians, in EMG pattern classification by modeling the EMG feature vectors using GMMs~\citep{Chan2003-hz}.
Huang \textit{et al.} also demonstrated that GMMs could classify EMG patterns with equal or better accuracy than that of an MLP~\citep{Huang2005-xi}.
LDA is likely the most popular linear generative classifier in EMG classification, consisting of a Gaussian distribution with a shared covariance matrix across all classes.
Although LDA is a linear classifier with a simple structure, it has been shown to be comparable in terms of accuracy to the SVM approach in some cases~\citep{Dellacasa_Bellingegni2017-ke}.

However, recent studies have suggested that EMG signals often follow a distribution that is more heavily tailed than a Gaussian distribution~\citep{Nazarpour2013-ss,Messaoudi2017-nj,Furui2019-ho}.
Such non-Gaussian properties of EMG signals are believed to be due to the variation in EMG variances depending on muscle activity~\citep{Milner-Brown1975-wy,Mcgill2004-ep,Furui2019-ho,Furui2019-os}.
Therefore, Gaussian distribution-based classification may not be sufficient as a generative approach.
To develop a classifier that reflects the stochastic nature of EMG signals, such varying non-Gaussianity needs to be considered.
Meanwhile, the authors previously proposed a stochastic model of EMG signals based on a scale mixture distribution assuming the EMG variance to be a random variable~\citep{Furui2019-ho}.
This model can represent changes in non-Gaussianity associated with muscle activity, thereby explaining both non-Gaussian and Gaussian distributions within a unified scheme.
If we can construct a classifier that incorporates the scale mixture model, a novel recognition framework that considers the stochastic properties of EMG can be established, thereby enhancing classification performance.

To this end, this paper aims to develop a classification method based on a scale mixture-based generative model of EMG signals and verify its effectiveness in EMG pattern recognition through experiments.
The stochastic model included in the proposed method is extended by making the distributions multidimensional and performing finite mixing to accommodate EMG feature vectors obtained from multiple electrodes.
In the framework of the scale mixture model, the scale parameter (e.g., variance or covariance) of the observed variable is considered as a latent random variable; therefore, the proposed method is capable of creating inference considering the variability superimposed on the scale parameter.
The proposed model is trained using variational Bayesian learning, enabling the automatic determination of the number of mixture components by pruning out the redundant components.
Furthermore, a maximum mutual information-based hyperparameter determination method is introduced.
We experimentally evaluate the validity of the proposed method through simulations, EMG analysis experiments, and classification experiments using public datasets.

\section{Proposed Classification Method}

\begin{figure}[t]
  \centering
	\includegraphics[width=0.75\hsize]{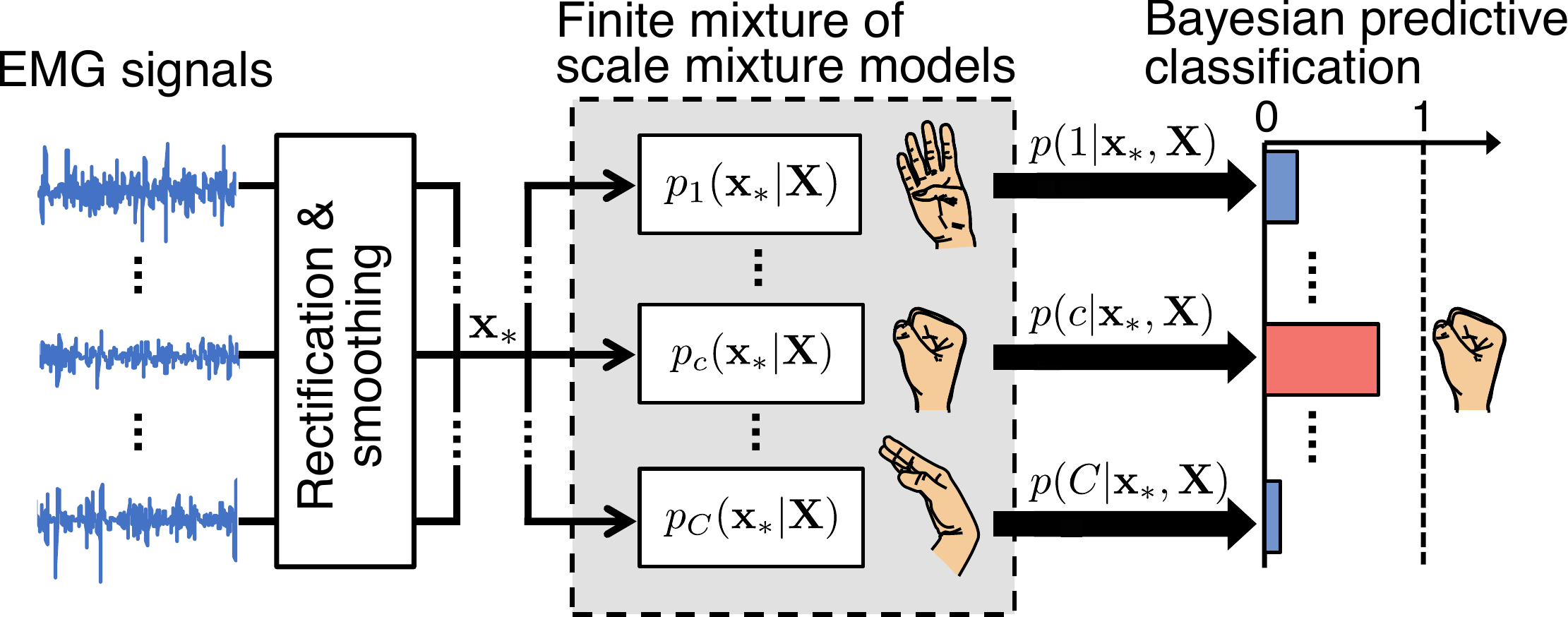}
	\caption{Overview of the proposed classification method incorporating scale mixture-based generative models.
	The measured EMG signals are first processed by rectification and smoothing and then input into the trained stochastic model corresponding to each motion $c$.
	Thereafter, the posterior distribution of class $c$ is calculated based on the Bayesian predictive classification.
	Finally, the class with the highest posterior probability is determined as the classified motion.
	}
	\label{fig:classifier}
\end{figure}

Fig.~\ref{fig:classifier} provides an overview of the proposed classification method.
This method incorporates stochastic models based on the finite mixture of scale mixture distributions, enabling classification by considering the variability in the variance.
Pattern recognition is achieved by inputting the EMG feature vectors into the stochastic model corresponding to each class and computing the class posterior distribution using a Bayesian predictive distribution.
The parameters of the proposed method, including the hyperparameters, are optimized based on variational Bayesian learning and mutual information maximization.

%
%

\subsection{Finite Mixture of Scale Mixture Models}
The scale mixture model of surface EMG signals is a stochastic model for raw EMG signals recorded from a pair of electrodes~\citep{Furui2019-ho}.
However, EMG pattern classification typically involves the use of EMG signals from multichannel electrodes.
In addition, the EMG signals utilized in such applications are processed by feature extraction techniques, such as rectification and smoothing, and their distributions may therefore exhibit skewness and multimodality.
Thus, we expand the previous scale mixture model to consider the multidimensionality and flexibility of the processed EMG signals and develop a finite mixture of scale mixture models.

\begin{figure}[t]
	\centering
  \includegraphics[width=0.6\hsize]{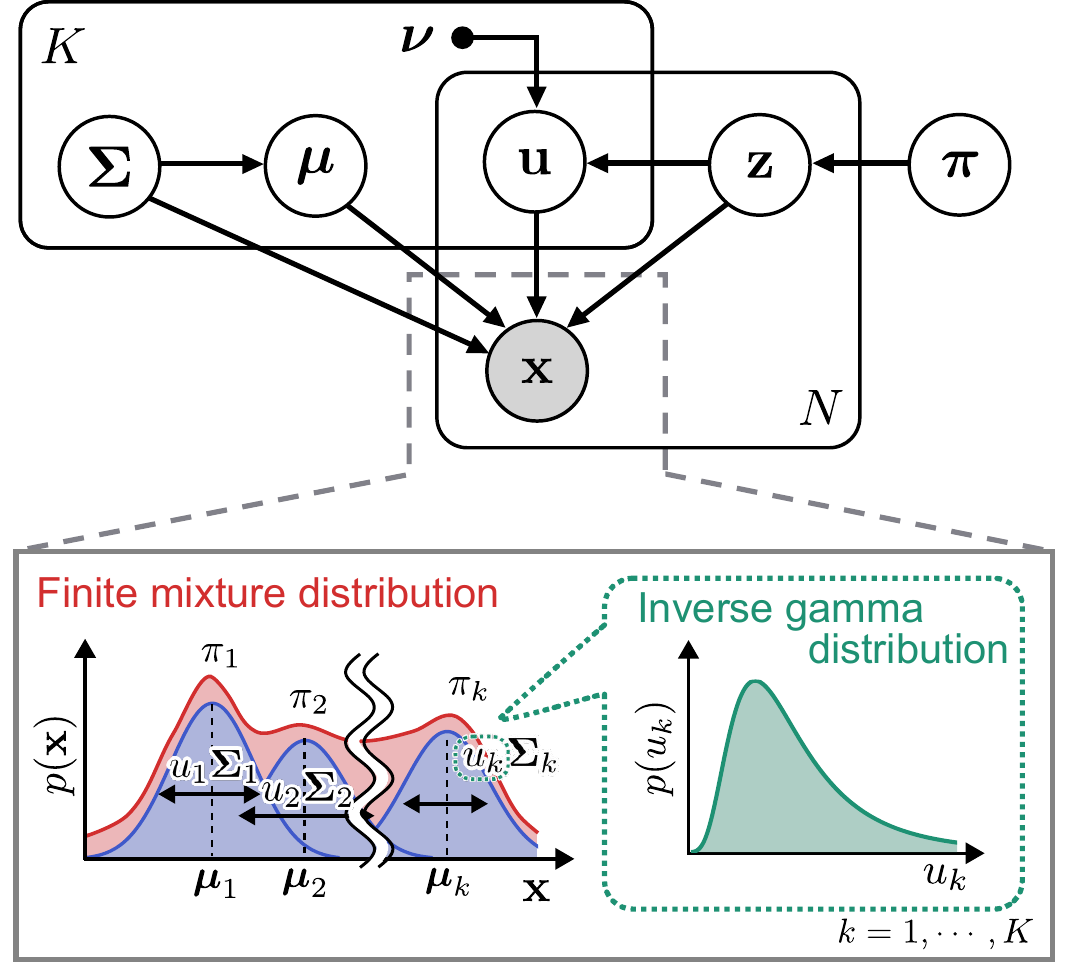}
	\caption{Graphical representation of the finite mixture of multivariate scale mixture models. 
	The diagram depicts a set of $N$ data points, with their corresponding latent points.
	In the model, an EMG feature vector $\mathbf{x}$ is handled as a random variable that follows a Gaussian mixture distribution. 
	A covariance matrix of each component in the Gaussian mixture distribution is weighted by the scale parameter $u_k$, which is a latent random variable following an inverse gamma distribution.}
	\label{fig:fmsmm_model}
\end{figure}

Fig.~\ref{fig:fmsmm_model} shows an overview of the finite mixture of multivariate scale mixture models.
%
%
The $D$-dimensional EMG feature vector $\mathbf{x}\in \mathbb{R}^{D}$ is handled as a random variable that follows a multivariate Gaussian mixture distribution, achieving flexible modeling of the distribution shape.
Each covariance matrix $\boldsymbol{\Sigma}_k \in \mathbb{R}^{D \times D}$ ($k = 1, \ldots, K; K$ is the number of components) in the Gaussian mixture distribution is weighted by a scale parameter $u_k \in \mathbb{R}^{+}$, which is a random variable following an inverse gamma distribution.
In the model, $u_k$ is interpreted as a latent variable because it is not directly observed.
The mixing condition is expressed by another latent variable $\mathbf{z} \in \mathbb{R}^K$, which corresponds to observations.
Accordingly, the model includes two latent variables, namely $u_k$ and $\mathbf{z}$.

For component $k$, the conditional distribution of the EMG feature vector $\mathbf{x}$ given $u_k$ is expressed via the following multivariate Gaussian distribution:
\begin{align}
    \mathcal{N}(\mathbf{x}|\boldsymbol{\mu}_k, u_k \mathbf{\Sigma}_k)
    & = (2 \pi)^{-\frac{D}{2}} |u_k \mathbf{\Sigma}_k|^{-\frac{1}{2}} \exp \left[-\frac{u_k^{-1}}{2}\Delta_k^2 \right],
	\label{eq:conditinal_distribution}
\end{align}
where $\boldsymbol{\mu}_k \in \mathbb{R}^D$ is the mean vector of component $k$ and $\Delta_k^2$ is the squared Mahalanobis distance defined by
\begin{equation}
	\Delta_k^2 = (\mathbf{x} - \boldsymbol{\mu}_k)^\mathrm{T} \mathbf{\Sigma}_k^{-1} (\mathbf{x} - \boldsymbol{\mu}_k).
\end{equation}
The distribution of $u_k$ is assumed to be the inverse gamma distribution:
\begin{align}
	\mathrm{IG}\left(u_k \middle| \frac{\nu_k}{2},\frac{\nu_k}{2} \right) 
	&= \frac{(\frac{\nu_k}{2})^{\frac{\nu_k}{2}}}{\Gamma (\frac{\nu_k}{2})}(u_k)^{-\frac{\nu_k}{2}-1} {\exp} \left[-\frac{\nu_k}{2 u_k}\right].
	\label{eq:scale_dist}
\end{align}
Parameter $\nu_k \in \mathbb{R}^{+}$ represents the degrees of freedom.
The multivariate scale mixture model for component $k$ can then be derived from (\ref{eq:conditinal_distribution}) and (\ref{eq:scale_dist}).
 \begin{align}
	p(\mathbf{x}| \boldsymbol{\mu}_k, &\mathbf{\Sigma}_k, \nu_k) \nonumber \\
	&= \int \mathcal{N}(\mathbf{x}|\boldsymbol{\mu}_k, u_k \mathbf{\Sigma}_k) \mathrm{IG}\left(u_k \middle| \frac{\nu_k}{2},\frac{\nu_k}{2} \right) \mathrm{d}u_k \label{eq:marginal_x} \\
	&= \frac{\Gamma(\frac{\nu_k+D}{2})}{\Gamma(\frac{\nu_k}{2})} \frac{|\mathbf{\Sigma}_k|^{-\frac{1}{2}}}{\left(\pi \nu_k\right)^{\frac{D}{2}}} \left(1+\Delta_k^2 \right)^{-\frac{\nu_k+D}{2}}.
\end{align}

To assign observations to each component, latent variable $\mathbf{z} = \{z_k \}$ based on 1-of-$K$ representation is introduced.
Therefore, the distribution of $\mathbf{z}$ can be expressed via the following categorical distribution:
\begin{equation}
	p(\mathbf{z}|\boldsymbol{\pi}) = \mathrm{Cat}(\mathbf{z}|\boldsymbol{\pi}) = \prod_{k=1}^{K} \pi_k^{z_k},
\end{equation}
where $\boldsymbol{\pi} = \{\pi_k \}$ is the mixing coefficients ($\pi_k \in [0, 1]$ and $\sum_{k=1}^{K} \pi_k = 1$).
Let us consider the conditional distributions of (\ref{eq:conditinal_distribution}) and (\ref{eq:scale_dist}) given a particular value of $\mathbf{z}$:
\begin{equation}
	p(\mathbf{x}|\mathbf{z},\mathbf{u}, \boldsymbol{\mu}, \mathbf{\Sigma}) = \prod_{k=1}^{K} \mathcal{N}(\mathbf{x}|\boldsymbol{\mu}_k, u_k \mathbf{\Sigma}_k)^{z_k},
\end{equation}
\begin{equation}
	p(\mathbf{u}|\mathbf{z}, \boldsymbol{\nu}) = \prod_{k=1}^{K} \mathrm{IG}\left(u_k \middle| \frac{\nu_k}{2},\frac{\nu_k}{2} \right)^{z_k},
\end{equation}
where $\mathbf{u} = \{u_k\}$.
The marginal distribution of $\mathbf{x}$ is then obtained by summing the joint distribution over all possible states of $\mathbf{z}$:
\begin{align}
	p(\mathbf{x}|\boldsymbol{\theta}, \boldsymbol{\nu}) &= \sum_{\mathbf{z}} p(\mathbf{z}|\boldsymbol{\pi}) \int p(\mathbf{x}|\mathbf{z},\mathbf{u}, \boldsymbol{\mu}, \mathbf{\Sigma}) p(\mathbf{u}|\mathbf{z}, \boldsymbol{\nu}) \mathrm{d}\mathbf{u} \\
	&= \sum_{k=1}^{K} \pi_k p(\mathbf{x}| \boldsymbol{\mu}_k, \mathbf{\Sigma}_k, \nu_k),
\end{align}
where $\boldsymbol{\theta} = \{\pi_k, \boldsymbol{\mu}_k, \mathbf{\Sigma}_k\}$.
Thus, the marginal distribution of $\mathbf{x}$ is given by a linear combination of the multivariate scale mixture models.

In Bayesian treatment, we additionally require priors over the model parameters, i.e., $\boldsymbol{\pi}$, $\boldsymbol{\mu}_k$, and $\mathbf{\Sigma}_k$.
To simplify the analysis, we introduce the corresponding conjugate prior distributions.
The prior on the mixing coefficients $\boldsymbol{\pi}$ is chosen to be a Dirichlet distribution 
\begin{equation}
	p(\boldsymbol{\pi}) = \mathrm{Dir}(\boldsymbol{\pi}|\boldsymbol{\alpha}_0) = \mathcal{C}(\boldsymbol{\alpha}_0) \prod_{k=1}^K \pi_k^{\alpha_0 - 1},
\end{equation}
where $\mathcal{C}(\cdot)$ is the normalization constant for the Dirichlet distribution given by
\begin{equation}
	\mathcal{C}(\boldsymbol{\alpha}) = \frac{\Gamma \left(\sum_{k=1}^K \alpha_k \right)}{\prod_{k=1}^K \Gamma(\alpha_k)}.
\end{equation}
Similarly, we introduce a Gaussian-inverse Wishart prior for the mean and covariance matrix of each component:
\begin{align}
	p(\boldsymbol{\mu}_k, \mathbf{\Sigma}_k) &= p(\boldsymbol{\mu}_k|\mathbf{\Sigma}_k)p(\mathbf{\Sigma}_k) \nonumber \\
										 &= \mathcal{N}(\boldsymbol{\mu}_k|\mathbf{m}_0, \beta_0^{-1} \mathbf{\Sigma}_k) \mathcal{IW}(\mathbf{\Sigma}_k|\mathbf{W}_0, \eta_0).
	\label{eq:prior_NIW}
\end{align}
The inverse Wishart part in (\ref{eq:prior_NIW}) is defined as
\begin{align}
	\mathcal{IW}&(\mathbf{\Sigma}_k|\mathbf{W}_0, \eta_0)\nonumber \\
    		&= \mathcal{B}(\mathbf{W}_0, \eta_0) |\mathbf{W}_0|^{-\frac{\eta_0+D+1}{2}}
			\exp \left[ -\frac{\mathrm{tr}(\mathbf{\Sigma}_k \mathbf{W}_0^{-1})}{2} \right],
\end{align}
where $\mathcal{B}(\cdot)$ is the normalization constant for the inverse Wishart distribution given by 
\begin{equation}
	\mathcal{B}(\mathbf{W}_k, \eta_k) = |\mathbf{W}_k|^{\frac{\eta_k}{2}} \left[ 2^{\frac{\eta_k D}{2}} \Gamma_D \left(\frac{\eta_k}{2} \right) \right]^{-1}.
\end{equation}
As there is no conjugate prior for $\nu_k$, it is considered as a non-random variable, indicating that no prior is imposed on it.

\subsection{Bayesian Predictive Classification}

The proposed model is built for each class using the given set of training data $\mathbf{X}$.
To classify the novel input vector $\mathbf{x}_*$ into one of the given $C$ classes (motions), the posterior distribution of class $c \in \{1, \ldots, C\}$ is calculated as
\begin{align}
	p(c|\mathbf{x}_*, \mathbf{X}) &= \frac{p_c(\mathbf{x}_*|\mathbf{X})p(c)}{\sum_{c'=1}^C p_{c'}(\mathbf{x}_*|\mathbf{X})p(c')}, \\
	p_c(\mathbf{x}_*|\mathbf{X}) &= \int p(\mathbf{x}_*|\mathbf{\Theta}_c) p(\mathbf{\Theta}_c|\mathbf{X}) \mathrm{d}\mathbf{\Theta}_c,
\end{align}
where $p_c(\mathbf{x}_*|\mathbf{X})$ is the posterior predictive distribution of $\mathbf{x}_*$ for class $c$, defined using the model built with the training data $\mathbf{X}$, $p(c)$ is the prior distribution of $c$, and $\mathbf{\Theta}_c = \{z_{ck}, u_{ck}, \pi_{ck}, \boldsymbol{\mu}_{ck}, \mathbf{\Sigma}_{ck}\}$ denotes the set of all unobserved random variables of the proposed model of $c$.
For a given $\mathbf{x}_*$, the class with the maximum posterior probability $p(c|\mathbf{x}_*, \mathbf{X})$ is determined as the classified motion.
The next subsection outlines the learning algorithm of the proposed model and the derivation of the predictive distribution based on variational Bayesian inference.

\subsection{Variational Bayesian Inference}
\subsubsection{Learning algorithm}

A set of $D$-dimensional EMG feature vectors and class labels $\{(\mathbf{x}_1, c_1),$ $\ldots,$ $(\mathbf{x}_N, c_N)\}$ is given as a training set to train the model corresponding to each class $c$.
The aim of Bayesian learning is to compute the posterior distribution and accompanying model evidence $p_c(\mathbf{X})$ for each class, where $\mathbf{X} = \{\mathbf{x}_1, \ldots, \mathbf{x}_N\}$.
However, the marginalization over the latent variables $\{z_{nck}, u_{nck}\}$ and parameters $\boldsymbol{\theta}_c$ to calculate the model evidence is intractable.
Therefore, we approximate the posterior distribution based on the variational inference framework and maximize the evidence lower bound to realize model training.

The logarithm of the model evidence to be maximized can be decomposed as follows~\citep{Bishop2006-th}:
\begin{equation}
	\ln p_c(\mathbf{X}) = \mathcal{L}_c + \mathrm{KL}[q(\mathbf{\Theta}_c)||p(\mathbf{\Theta}_c|\mathbf{X})],
\label{eq:evidence}
\end{equation}
where $\mathcal{L}_c$ is the evidence lower bound defined by
\begin{equation}
	\mathcal{L}_c = \int q(\mathbf{\Theta}_c) \ln \left\{\frac{p(\mathbf{X},\mathbf{\Theta}_c)}{q(\mathbf{\Theta}_c)} \right\} \mathrm{d}\mathbf{\Theta}_c,
\end{equation}
where $q(\mathbf{\Theta}_c)$ denotes a variational posterior distribution approximating the true posterior $p(\mathbf{\Theta}_c|\mathbf{X})$. 
The term $\mathrm{KL}[q(\mathbf{\Theta}_c)||p(\mathbf{\Theta}_c|\mathbf{X})]$ is the Kullback-Leibler (KL) divergence between the approximate and true posteriors.
According to (\ref{eq:evidence}), minimizing the KL divergence between $p(\mathbf{\Theta}_c|\mathbf{X})$ and $q(\mathbf{\Theta}_c)$ maximizes the evidence lower bound $\mathcal{L}_c$, which can be achieved by optimizing $q(\mathbf{\Theta}_c)$ based on the variational inference framework.

In variational Bayesian inference, the variational distribution $q(\mathbf{\Theta}_c)$ is assumed to be factorized over some partition of unobserved random variables, i.e., $q(\mathbf{\Theta}_c) = \prod_m q(\mathbf{\Theta}_{cm})$, enabling an efficient solution.
A general expression for the optimal solution is~\citep{Bishop2006-th}:
\begin{equation}
	\ln q(\mathbf{\Theta}_{cm}) = \langle \ln p(\mathbf{X}, \mathbf{\Theta}_c) \rangle_{\mathbf{\Theta}_{c \backslash m}} + \mathrm{const},
\end{equation}
where $\langle \cdot \rangle_{\mathbf{\Theta}_{c \backslash m}}$ denotes the expectation with respect to the $q$ distributions over all unobserved variables excluding the $m$-th group.
In this paper, we assume that the variational distribution is only factorized over the latent variables and parameters:
\begin{align}
	q(\mathbf{\Theta}_c) = q(\mathbf{Z}_c, \mathbf{U}_c) q(\boldsymbol{\pi}_c, \boldsymbol{\mu}_c, \mathbf{\Sigma}_c).
\end{align}
Based on this factorization, we optimize the factors of the latent variables and parameters in order.

Let us consider the derivation of the update equation for $q(\mathbf{Z}_c, \mathbf{U}_c)$.
The logarithm of the optimal factor is given by
\begin{align}
	\ln q(\mathbf{z}_{nc}, \mathbf{u}_{nc}) &= \langle \ln p(\mathbf{x}_n|\mathbf{z}_{nc}, \mathbf{u}_{nc}, \boldsymbol{\mu}_c, \mathbf{\Sigma}_c) \rangle_{\boldsymbol{\mu}_c, \mathbf{\Sigma}_c} + \ln p(\mathbf{u}_{nc}|\mathbf{z}_{nc}) \nonumber \\
	&\quad + \langle \ln p(\mathbf{z}_{nc}|\boldsymbol{\pi}_c) \rangle_{\boldsymbol{\pi}_c} + \mathrm{const} \nonumber \\
	&= \sum_{k=1}^{K_c} z_{nck} \ln \rho_{nck} + \mathrm{const},
\end{align}
where we define
\begin{align}
	\ln \rho_{nck} &= \langle \ln \pi_{ck} \rangle -\frac{D}{2} \ln(2\pi) - \frac{D}{2} \ln u_{nck} -\frac{1}{2} \langle {\ln |\mathbf{\Sigma}_{ck}|} \rangle \nonumber \\
	&\quad -\frac{1}{2} u_{nck}^{-1} \left\langle \Delta_{nck}^2 \right\rangle + \frac{\nu_{ck}}{2} \ln \frac{\nu_{ck}}{2} - \ln \Gamma \left(\frac{\nu_{ck}}{2} \right) \nonumber \\
	&\quad   - \left(\frac{\nu_{ck}}{2} + 1 \right) \ln u_{nck} - \frac{\nu_{ck}}{2} u_{nck}^{-1}.
	\label{eq:ln_rho}
\end{align}
First, the variational posterior distribution $q(\mathbf{z}_{nc})$ can be determined by marginalizing $q(\mathbf{z}_{nc}, \mathbf{u}_{nc})$ over $\mathbf{u}_{nc}$, resulting in the form of a categorical distribution.
\begin{align}
	q(\mathbf{z}_{nc}) = \mathrm{Cat}(\mathbf{z}_{nc}|\mathbf{r}_{nc}),
	\label{eq:posterior_cat}
\end{align}
where $\mathbf{r}_{nc} = \{r_{nck} \}$ play the role of responsibilities and are given by
\begin{align}
	r_{nck} = \frac{\int \rho_{nck}\, \mathrm{d}u_{nck}}{\sum_{k'=1}^K \int \rho_{nck'}\, \mathrm{d}u_{nck'}}.
	\label{eq:responsibility}
\end{align}
Next, using the product rule, the factor $q(\mathbf{z}_{nc}, \mathbf{u}_{nc})$ can be written in the form $q(\mathbf{z}_{nc}, \mathbf{u}_{nc}) = q(\mathbf{u}_{nc}|\mathbf{z}_{nc})q(\mathbf{z}_{nc})$.
Accordingly, the variational posterior distribution on the scale parameter, $q(\mathbf{u}_{nc}|\mathbf{z}_{nc} = 1)$, has the form of an inverse gamma distribution:
\begin{align}
	q(u_{nck}|z_{nck} = 1) = \mathrm{IG}(u_{nck} | a_{nck}, b_{nck}),
	\label{eq:posterior_IG}
\end{align}
where
\begin{align}
	a_{nck} &= \frac{\nu_{ck} + D}{2}, \\
	b_{nck} &= \frac{1}{2} \langle \Delta_{nck}^2 \rangle + \frac{\nu_{ck}}{2}.
\end{align}

Now, let us consider the factor $q(\boldsymbol{\pi}_c, \boldsymbol{\mu}_c, \mathbf{\Sigma}_c)$ in the variational posterior distribution.
To complete the calculations of the variational posterior distributions over the model parameters, we first define some statistics for convenience:
\begin{align}
	N_{ck} &= \sum_{n} \langle z_{nck} \rangle, \label{eq:start_stats} \\
	\omega_{ck} &= \sum_{n} \langle z_{nck} \rangle \langle u^{-1}_{nck} \rangle, \\
	\overline{\mathbf{x}}_{ck} &= \omega_{ck}^{-1} \sum_{n} \langle z_{nck} \rangle \langle u^{-1}_{nck} \rangle \mathbf{x}_n, \\
	\mathbf{S}_{ck} &= \omega_{ck}^{-1} \sum_{n} \langle z_{nck} \rangle \langle u^{-1}_{nck} \rangle (\mathbf{x}_n - \overline{\mathbf{x}}_{ck})(\mathbf{x}_n - \overline{\mathbf{x}}_{ck})^\mathrm{T}, \label{eq:end_stats}
\end{align}
The logarithm of the optimal factor is given by
\begin{align}
	\ln q(\boldsymbol{\pi}_c, \boldsymbol{\mu}_c, \mathbf{\Sigma}_c) &= \langle \ln p(\mathbf{X}|\mathbf{Z}_c,\mathbf{U}_c,\boldsymbol{\mu}_c, \mathbf{\Sigma}_c) \rangle_{\mathbf{Z}_c,\mathbf{U}_c} + \langle \ln p(\mathbf{Z}_c|\boldsymbol{\pi}_c) \rangle_{\mathbf{Z}_c} \nonumber \\
	& \quad + \ln p(\boldsymbol{\pi}_c) + \ln p(\boldsymbol{\mu}_c, \mathbf{\Sigma}_c) + \mathrm{const}.
	\label{eq:vb_params}
\end{align}
By extracting the terms that depend on $\boldsymbol{\pi}_c$ from (\ref{eq:vb_params}), the variational posterior distribution $q(\boldsymbol{\pi}_c)$ becomes a Dirichlet distribution:
\begin{align}
	q(\boldsymbol{\pi}_c) = \mathrm{Dir}(\boldsymbol{\pi}_c|\boldsymbol{\alpha}_c),
	\label{eq:posterior_dir}
\end{align}
where $\boldsymbol{\alpha}_c = \{\alpha_{ck} \}$ is defined as
\begin{align}
	\alpha_{ck} = \alpha_0 + N_{ck}.
\end{align}
Similarly, the variational posterior distribution $q(\boldsymbol{\mu}_c, \mathbf{\Sigma}_c)$ is given by
\begin{align}
	q(\boldsymbol{\mu}_c, \mathbf{\Sigma}_c) = \mathcal{N}(\boldsymbol{\mu}_{ck} | \mathbf{m}_{ck}, \beta_{ck}^{-1} \mathbf{\Sigma}_{ck}) \mathcal{IW}(\mathbf{\Sigma}_{ck} | \mathbf{W}_{ck}, \eta_{ck}),
	\label{eq:posterior_NIW}
\end{align}
where we define
\begin{align}
	\beta_{ck} &= \beta_0 + \omega_{ck},\\
	\mathbf{m}_{ck} &= \frac{1}{\beta_{ck}}(\omega_{ck} \overline{\mathbf{x}}_{ck} + \beta_0 \mathbf{m}_0),\\
	\mathbf{W}_{ck} &= \mathbf{W}_0 + \omega_{ck} \mathbf{S}_{ck} + \frac{\beta_0 \omega_{ck}}{\beta_{ck}} (\overline{\mathbf{x}}_{ck} - \mathbf{m}_0)(\overline{\mathbf{x}}_{ck} - \mathbf{m}_0)^{\mathrm{T}}, \\
	\eta_{ck} &= \eta_0 + N_{ck}.
\end{align}

There is no closed-form update equation for the degrees of freedom $\nu_{ck}$, as it is not randomized due to the missing conjugate prior.
When a fully generative approach is adopted, the most straightforward means of determining $\nu_{ck}$ is to point-estimate its value by maximizing the log-marginal likelihood in each iteration of variational inference.
However, our previous study demonstrated that the classification accuracy for novel data decreases significantly when $\nu_{ck}$ is optimized for each class in a maximum likelihood manner~\citep{Furui2018-xq}.
This problem can be avoided by fixing $\nu_{ck}$ to a certain value $\widehat{\nu}$ and sharing it with all the classes and components.
Therefore, instead of maximizing the marginal likelihood for each class, this paper introduces a discriminative method of determining $\widehat{\nu}$ based on maximum mutual information estimation, thereby improving the generalization ability.
Section 2.4 presents the details of the method of determining $\nu_{ck} = \widehat{\nu}$.
Henceforth, when we give a general description of the degrees-of-freedom parameter of the proposed model, we simply refer to it as $\nu$ without specifying the component or class.

The expectation required for the update equation of each variational posterior distribution can be calculated as follows:
\begin{align}
	\langle z_{nck} \rangle &= r_{nck}, \label{eq:e_z}\\
	\langle u_{nck}^{-1} \rangle &= \frac{a_{nck}}{b_{nck}}, \label{eq:e_inverse_scale}\\
	\langle \ln u_{nck} \rangle &= \ln b_{nck} - \psi(a_{nck}), \label{eq:e_log_scale}\\
	\ln \widetilde{\Sigma}_{ck} &\triangleq  \langle \ln |\mathbf{\Sigma}_{ck}| \rangle \nonumber \\
	&= - \sum_{d=1}^D \psi \left(\frac{\eta_{ck}\! +\! 1\! - d}{2} \right) - D \ln 2 + \ln |\mathbf{W}_{ck}|, \label{eq:moment_Sigma}\\
	\langle \Delta_{nck}^2 \rangle &= D \beta_{ck}^{-1} + \eta_{ck} (\mathbf{x}_n - \mathbf{m}_{ck})^{\mathrm{T}}\mathbf{W}_{ck}^{-1}(\mathbf{x}_n - \mathbf{m}_{ck}), \label{eq:moment_delta}\\
	\ln \widetilde{\pi}_{ck} &\triangleq \langle \ln \pi_{ck} \rangle = \psi(\alpha_{ck}) - \psi \left(\widehat{\alpha}_c \right) \label{eq:moment_pi},
\end{align}
where $\widehat{\alpha}_c  = \sum_{k} \alpha_{ck}$.
By substituting (\ref{eq:moment_Sigma})--(\ref{eq:moment_pi}) into (\ref{eq:ln_rho}) and calculating the integral in (\ref{eq:responsibility}), we obtain the following result for the responsibilities
\begin{align}
	r_{nck} &\propto \mathbbm{1}(c_n = c) \frac{\Gamma(\frac{\nu_{ck}+D}{2})}{\Gamma(\frac{\nu_{ck}}{2}) (\pi \nu_{ck})^{\frac{D}{2}}}  \widetilde{\pi}_{ck} \widetilde{\Sigma}_{ck}^{-\frac{1}{2}} \left\{ 1 + \frac{1}{\nu_{ck}} \langle \Delta_{nck}^2 \rangle \right\}^{-\frac{\nu_{ck} + D}{2}}.
	\label{eq:resp}
\end{align}
where $\mathbbm{1}(e)$ denotes the indicator function, which is 1 when $e$ is true and 0 otherwise.
This function is used to assign only the data points corresponding to each class label to $r_{nck}$.

From the above, the optimization of variational posterior distributions for each class is achieved by repeating the following two steps: 1) update the posterior distributions of the latent variables using those of the current parameters, and 2) update the posterior distributions of the parameters using those of the current latent variables.
The evidence lower bound for class $c$ is iteratively maximized in each of these steps, and is calculated as follows:
	\begin{align}
		\mathcal{L}_c &= \langle \ln p(\mathbf{X}|\mathbf{Z}_c, \mathbf{U}_c, \boldsymbol{\mu}_c, \mathbf{\Sigma}_c) \rangle_{\mathbf{\Theta}_c} + \langle \ln p(\mathbf{Z}_c, \mathbf{U}_c|\boldsymbol{\pi}_c) \rangle_{\mathbf{\Theta}_c} \nonumber \\
		&= \langle \ln p(\boldsymbol{\theta}_c) \rangle_{\boldsymbol{\theta}_c} - \langle \ln q(\mathbf{Z}_c, \mathbf{U}_c) \rangle_{\mathbf{Z}_c, \mathbf{U}_c} - \langle \ln q(\boldsymbol{\theta}_c) \rangle_{\boldsymbol{\theta}_c}.
		\label{eq:ELBO}
	\end{align}
During the learning process, the $k$-th component is pruned out from the model when $\langle z_{nck} \rangle$ approaches 0.
Because of the nature of the Dirichlet distribution set as the prior distribution of the mixing coefficients, the smaller the value of the prior parameter $\alpha_0$, the more aggressive the pruning of the mixing components.
The training procedure of the proposed method is summarized in Algorithm~\ref{al:learning}.

\begin{algorithm}[t]
	\caption{Learning algorithm of the proposed model}
	\label{al:learning}
	
	\SetAlgoLined
		\SetKwInOut{Input}{Input}
		\Input{Training data $\mathbf{X}$, corresponding class label $\mathbf{c}$, fixed $\widehat{\nu}$, and prior parameters $\alpha_0$, $\beta_0$, $\mathbf{m}_0$, $\mathbf{W}_0$, $\eta_0$}
		\SetKwInOut{Output}{Output}
		\Output{Variational posterior distributions $q(\boldsymbol{\pi}_c)$ and $q(\boldsymbol{\mu}_c, \mathbf{\Sigma}_c)$}
		Initialize the posterior distributions\;
		\For{each class $c \in \{1,\ldots,C \}$}{
			Set $\nu_{ck} \leftarrow  \widehat{\nu}$\;
			\While{$\mathcal{L}_c$ have not converged}{
				Calculate $\ln \widetilde{\Sigma}_{ck}$, $\langle \Delta_{nck}^2 \rangle$, $\ln \widetilde{\pi}_{ck}$ using (\ref{eq:moment_Sigma})--(\ref{eq:moment_pi})\;
				Calculate $r_{nck}$ using (\ref{eq:responsibility}) and (\ref{eq:resp})\;
				Update $q(\mathbf{Z}_c)$ and $q(\mathbf{U}_c)$ using (\ref{eq:posterior_cat}) and (\ref{eq:posterior_IG})\;
				Evaluate $N_{ck}$, $w_{ck}$, $\overline{\mathbf{x}}_{ck}$, $\mathbf{S}_{ck}$ using (\ref{eq:start_stats})--(\ref{eq:end_stats})\;
				Calculate $\langle z_{nck} \rangle$, $\langle u_{nck}^{-1} \rangle$, $\langle \ln u_{nck} \rangle$ using (\ref{eq:e_z})--(\ref{eq:e_log_scale})\;
				Update $q(\boldsymbol{\pi}_c)$ and $q(\boldsymbol{\mu}_c, \mathbf{\Sigma}_c)$ using (\ref{eq:posterior_dir}) and (\ref{eq:posterior_NIW})\;
				Evaluate $\mathcal{L}_c$ using (\ref{eq:ELBO})\;
			}
		}
\end{algorithm}

\subsubsection{Posterior predictive distribution}
The posterior predictive distribution of novel input $\mathbf{x}_*$ given $\mathbf{X}$ is calculated by marginalizing the posterior distribution with respect to all the unobserved variables of no interest:
\begin{align}
	p_c(\mathbf{x}_*|\mathbf{X}) &= \int p(\mathbf{x}_*|\mathbf{\Theta}_c) p(\mathbf{\Theta}_c|\mathbf{X}) \mathrm{d}\mathbf{\Theta}_c \nonumber \\
	& = \sum_{\mathbf{z}_{*c}} \iiiint p(\mathbf{x}_{*}|\mathbf{z}_{*c}, \mathbf{u}_{*c}, \boldsymbol{\mu}_c, \mathbf{\Sigma}_c) p(\mathbf{u}_{*c}|\mathbf{z}_{*c}) \nonumber  \\
	&\qquad \quad \times p(\mathbf{z}_{*c}|\boldsymbol{\pi}_c) p(\boldsymbol{\pi}_c, \boldsymbol{\mu}_c, \mathbf{\Sigma}_c|\mathbf{X})\,  \mathrm{d}\mathbf{u}_{*c}\, \mathrm{d}\boldsymbol{\pi}_c\, \mathrm{d}\boldsymbol{\mu}_c\, \mathrm{d}\mathbf{\Sigma}_c \nonumber \\
	&\simeq \sum_{k=1}^{K_c} \iiiint  \pi_{ck} \mathcal{N}(\mathbf{x}_*|\boldsymbol{\mu}_{ck}, u_{*ck} \mathbf{\Sigma}_{ck}) \mathrm{IG}\left(u_{*ck} \middle| \frac{\nu_{ck}}{2}, \frac{\nu_{ck}}{2} \right) \nonumber \\
	&\qquad \qquad \times q(\boldsymbol{\pi}_c) q(\boldsymbol{\mu}_c, \mathbf{\Sigma}_c)\,  \mathrm{d}u_{*ck}\, \mathrm{d}\boldsymbol{\pi}_c\, \mathrm{d}\boldsymbol{\mu}_{ck}\, \mathrm{d}\mathbf{\Sigma}_{ck} \nonumber \\
	& = \sum_{k=1}^{K_c} \langle \boldsymbol{\pi}_c \rangle \iint p(\mathbf{x}_*|\boldsymbol{\mu}_{ck}, \mathbf{\Sigma}_{ck}, \nu_{ck}) \mathcal{N}(\boldsymbol{\mu}_{ck}|\mathbf{m}_{ck}, \beta_{ck}^{-1} \mathbf{\Sigma}_{ck}) \nonumber \\
	&\qquad \qquad \times \mathcal{IW}(\mathbf{\Sigma}_{ck}|\mathbf{W}_{ck}, \eta_{ck})\,  \mathrm{d}\boldsymbol{\mu}_{ck}\, \mathrm{d}\mathbf{\Sigma}_{ck}.
\end{align}
To obtain an analytical solution for the predictive distribution, we approximate $p_c(\mathbf{x}_*|\mathbf{X})$ as follows:
\begin{align}
	p_c(\mathbf{x}_*|\mathbf{X})
	& \simeq \sum_{k=1}^{K_c} \langle \boldsymbol{\pi}_c \rangle p(\mathbf{x}_*|\langle \boldsymbol{\mu}_{ck} \rangle, \langle \mathbf{\Sigma}_{ck} \rangle, \nu_{ck}).
\end{align}
Each posterior expectation can be calculated as $\langle \boldsymbol{\pi}_c \rangle = \alpha_{ck} / \widehat{\alpha}_c$, $\langle \boldsymbol{\mu}_{ck} \rangle = \mathbf{m}_{ck}$, and $\langle \mathbf{\Sigma}_{ck} \rangle = (\eta_{ck} - D - 1)^{-1} \mathbf{W}_{ck}$.

\subsection{Determination of the Degrees-of-Freedom Parameter Based on Maximum Mutual Information}

First, a training set $\{(\mathbf{x}_1, c_1), \ldots, (\mathbf{x}_N, c_N)\}$ is divided into $L$ subsets.
The $L - 1$ subsets are considered as a partial training set, and the remaining $l$-th subset $(l = 1, \ldots, L)$ is considered as the validation set.
Then, training based on the variational Bayesian inference with $\nu_{ck}$ fixed to $\nu^\mathrm{pre}$ is performed on the partial training set to obtain the variational posterior distributions for each class.
We also define the mutual information for the validation set as follows:
\begin{equation}
	\mathrm{I}[\mathbf{X}^{(l)}, \mathbf{c}^{(l)}] = \mathrm{H}[\mathbf{c}^{(l)}] - \mathrm{H}[\mathbf{c}^{(l)}|\mathbf{X}^{(l)}],
\end{equation}
where $\mathbf{X}^{(l)}$ and $\mathbf{c}^{(l)}$ are the validation data matrix and corresponding label vector with $c^{(l)}_n$ as its elements in the $l$-th subset, respectively, and they contain $N^{(l)}$ data points.
$\mathrm{H}[\mathbf{c}^{(l)}]$ and $\mathrm{H}[\mathbf{c}^{(l)}|\mathbf{X}^{(l)}]$ are the marginal entropy and conditional entropy, respectively.
The mutual information $\mathrm{I}[\mathbf{X}^{(l)}, \mathbf{c}^{(l)}]$ represents the mutual dependence between the data and corresponding classes.
Therefore, searching for a parameter that maximizes the mutual information implies learning the parameter in a discriminative manner.

Maximizing $\mathrm{I}[\mathbf{X}^{(l)}, \mathbf{c}^{(l)}]$ is equivalent to minimizing the conditional entropy $\mathrm{H}[\mathbf{c}^{(l)}|\mathbf{X}^{(l)}]$ because $\mathrm{H}[\mathbf{c}^{(l)}]$ is fixed.
Accordingly, we define the objective function for minimization as follows:
\begin{align}
	J(\nu, \mathbf{X}^{(l)}, \mathbf{c}^{(l)}) &= \mathrm{H}[\mathbf{c}^{(l)}|\mathbf{X}^{(l)}] \nonumber\\ 
		&= -\langle \ln p(\mathbf{c}^{(l)}|\mathbf{X}^{(l)}, \mathbf{X}^{(\backslash l)}) \rangle_{\mathbf{c}^{(l)}} \nonumber\\
		&\simeq  -\frac{1}{N^{(l)}} \sum_{n=1}^{N^{(l)}} \ln p(c_n^{(l)}|\mathbf{x}^{(l)}_n, \mathbf{X}^{(\backslash l)}, \nu) \nonumber\\ 
		&= -\frac{1}{N^{(l)}} \sum_{n=1}^{N^{(l)}} \ln \frac{p_{c_n^{(l)}}(\mathbf{x}^{(l)}_n|\mathbf{X}^{(\backslash l)}, \nu) p(c^{(l)}_n)}{\sum_{c'=1}^C p_{c'}(\mathbf{x}^{(l)}_n|\mathbf{X}^{(\backslash l)}, \nu)p(c')}, \label{eq:JJ}
\end{align}
where $\mathbf{X}^{(\backslash l)}$ is the partial training data matrix excluding the $l$-th fold data and $p_c(\mathbf{x}^{(l)}_n|\mathbf{X}^{(\backslash l)}, \nu)$ is the posterior predictive distribution for class $c$ defined by the model trained with $\mathbf{X}^{(\backslash l)}$.
An optimal $\nu^{(l)}$ for the $l$-th subset is obtained by searching for a value of $\nu$ that minimizes the objective function, while keeping the variational posterior distributions that are unrelated to $\nu$ fixed, for the models obtained using $\nu^{\mathrm{pre}}$ and the partial training data $\mathbf{X}^{(\backslash l)}$.
\begin{equation}
	\nu^{(l)} = \argmin_\nu J(\nu, \mathbf{X}^{(l)}, \mathbf{c}^{(l)}).
\end{equation}
This procedure is performed for all $L$ folds, and the smallest value of the obtained $\{ \nu^{(l)} \}$ is set as the optimal $\widehat{\nu}$ for the entire training data $\mathbf{X}$.

\section{Experiments}

\subsection{Simulation}
To evaluate the classification characteristics of the proposed method, a simulation experiment was performed using two-dimensional synthetic data $\mathbf{x} = [x_1, x_2]^\mathrm{T}$ with two classes ($D = 2$ and $C = 2$).
The data corresponding to each class were artificially generated using the Gaussian distributions $p(\mathbf{x}|c=1) = \mathcal{N}(\mathbf{x}|[2.5, 2.5]^\mathrm{T}, 0.5 \mathbf{I})$ and $p(\mathbf{x}|c=2) = \mathcal{N}(\mathbf{x}|[5, 5]^\mathrm{T}, 0.5\mathbf{I})$, respectively, where $\mathbf{I} \in \mathbb{R}^2$ is the identity matrix.
We added outliers generated by the uniform distribution in the range $[0, 7]$ to the dataset of class 1 and evaluated the robustness against the outliers.

In the experiment, the number of training examples for each class was 100.
The number of outliers added to class~1 was set to 10\% of the data length, i.e., 10 data points.
After training the proposed classifier using the generated synthetic data, we visualized the decision boundary by inputting values in the ranges $0 \leq x_1 \leq 8$ and $0 \leq x_2 \leq 8$ as novel data and calculating the posterior class distribution.
To evaluate the effect of the degrees-of-freedom parameter $\nu$ in the proposed method, training and prediction of the decision boundaries were performed in two different settings: using $\nu$ values estimated for each class and component individually based on maximum likelihood and using $\nu = 5$ for all classes.
The corresponding decision boundaries were compared with those given by the GMM.
Both the proposed method and GMM had one component ($K_c = 1$).
The parameters of the prior distributions were set as follows: $\alpha_0 = 0.001$, $\beta_0 = 1$, and $\eta_0 = D + 1$.
$\mathbf{m}_0$ and $\mathbf{W}_0$ were set as the mean vector and covariance matrix calculated from the training data, respectively.

\subsection{EMG Analysis Experiments}
To evaluate the suitability of the proposed classification method for actual EMG data, an EMG analysis experiment was conducted.
Three healthy young adults (males, right-handed, aged 22--24 years old) were recruited in this experiment.
Four pairs of electrodes ($D = 4$) were attached to the surface of the skin in equal intervals near the elbow of the forearm, and the EMG signals were measured using a wireless measurement system (Delsys, Trigno; 16-bit A/D; sampling frequency: 2,000 Hz; bandwidth: 20--450 Hz).
During the experiment, the participants were instructed to perform 10 trials of six motions ($C = 6$: palmar extension, palmar flexion, supination, pronation, hand open, and hand grasp) for 7 s each in a seated posture with their right elbow fixed on the desk.
Experimental instructions were presented on a display in front of the participants, and no visual feedback of the classification results was presented.
The participants were informed of the aim of the study and provided written informed consent in advance.
The experiments were approved by the Hiroshima University Ethics Committee (Registration number: E-840).

Feature extraction was conducted through rectification and smoothing using a second-order Butterworth low-pass filter with a cut-off frequency of 2 Hz.
The data from the last 5 s of the 7 s of each motion, largely excluding the transient phase from rest to motion, were used for analysis.
This was done to evaluate the properties of the proposed method in as simple a situation as possible.
To evaluate the classification performance, 500 points randomly extracted from the data of one trial were used for the training of each motion, and all the data of the remaining trials were used for testing.
We verified all the combinations for each participant.
In the evaluation, we calculated the classification accuracy and number of components after training by varying the initial number of components and degrees of freedom respectively as $K_c = 1, 2, \ldots, 10$ and $\nu = 1, 2, \ldots, 10$.
For comparison, the classification accuracy was also calculated for the cases in which $\nu$ was estimated for each class and component (i.e., using $\nu_{ck}$) and the method of determining $\nu = \widehat{\nu}$ based on mutual information maximization was used.
The preset value $\nu^{\mathrm{pre}}$ was set to a sufficiently large value of 200.
The prior distribution settings were the same as in the simulation experiment.

\subsection{EMG Classification Experiments}
\subsubsection{Evaluation using various datasets}

To evaluate the capability of the proposed classifier quantitatively, we conducted a classification experiment using public EMG datasets.
Table~\ref{table:datasets} lists the characteristics of the six datasets.
\begin{table*}[]
	\centering
	\caption{Data description}
	\begin{adjustbox}{max width=\textwidth}
	\begin{tabular}{clllllll}
	\toprule 
	Dataset      & \# Motions	& \# Electrodes	& \# Participants	& \# All trials & \# Training trials & Training sample size	& Test sample size \\ 
	\midrule
	I	& 14	&	8	& 8		& 4		& 1	& 1,800 & 36,000 \\
	II	& 15	&	8	& 8		& 3		& 1	& 3,600 & 72,000 \\
	III	& 10	&	2	& 10	& 6		& 2	& 900 & 18,000 \\
	IV	& 7		&	8	& 44	& 8		& 2	& 230--691 & 4,613--13,836 \\
	V	& 7		&	8	& 44	& 6		& 2	& 230--691 & 4,613--13,836 \\
	VI	& 7		&	8	& 44	& 6		& 2	& 230--691 & 4,613--13,836 \\
	\bottomrule 
	\end{tabular}
\end{adjustbox}
	\label{table:datasets}
\end{table*}
Datasets I, II, and III were taken from \citep{Khushaba2013-mn}, \citep{Khushaba2012-ii}, and \citep{Khushaba2012-zo}, respectively, and can be downloaded from Dr. Khushaba's webpage\footnote{https://www.rami-khushaba.com/electromyogram-emg-repository.html}.
Datasets IV--VI were extracted from \textit{putEMG} dataset \citep{Kaczmarek2019-ns}, and is available on the webpage of Biomedical Engineering and Biocybernetics Team\footnote{https://biolab.put.poznan.pl/putemg-dataset/}.
\begin{itemize}
	\item \textbf{Dataset I}: Eight participants (aged 20--35 years) were recruited. 
	Each participant sat in an armchair, facing a personal computer with a steering wheel attached to the desk. 
	Each participant then performed 14 classes of finger movements ($C = 14$) over four trials ($T = 4$).
	EMG signals were recorded using eight-channel electrodes ($D = 8$) at 4,000 Hz and digitized using a 12-bit A/D converter.
	\item \textbf{Dataset II}: Eight participants (aged 20--35 years) were recruited.
	The participants sat in an armchair with their arm supported and fixed in one position.
	Each participant performed 15 classes of finger and hand movements ($C = 15$) over three trials ($T = 3$).
	The EMG signal recording settings were the same as those for Dataset I.
	\item \textbf{Dataset III}: Eight participants (aged 20--35 years) were recruited to perform the required fingers movements.
	The participants sat in an armchair with their arm supported and fixed in one position.
	EMG signals were recorded using two EMG sensors ($D = 2$) at 4,000 Hz and digitized using a 12-bit A/D converter.
	The participants performed 10 classes of individual and combined finger movements ($C = 10$) over six trials ($T = 6$).
	\item \textbf{Datasets IV--V}: Forty-four participants (aged 19--37 years) were recruited. 
	Three elastic bands with eight electrodes were placed around the forearm of each participant, and the participant performed seven active finger and hand movements ($C = 7$).
	In our experiment, only the middle band was used ($D = 8$).
	EMG signals were recorded at 5,120 Hz with a 12-bit A/D converter.
	Each participant performed three different tasks, which were used as different datasets in our experiment. Dataset IV included seven action blocks, with each block containing eight repetitions of each motion (called \textit{repeats\_long} in \citep{Kaczmarek2019-ns}). 
	Dataset V included seven action blocks, with each block containing six repetitions of each motion (called \textit{repeats\_short} in \citep{Kaczmarek2019-ns}). 
	Finally, Dataset VI included six action blocks, with each block being a subsequent execution of all the active gestures (called \textit{sequential} in \citep{Kaczmarek2019-ns}).
	The two sessions were conducted and we targeted only the first session.
\end{itemize}
Note that Datasets I--III contained data that were already segmented for each motion, implying that the transient states from rest to motion were excluded; in contrast, Datasets IV--VI were labeled for a series of data; thus, the data for each motion contained some transient states.
Feature extraction for the datasets was conducted by rectification and smoothing using a second-order Butterworth low-pass filter with a cut-off frequency of 2 Hz.

For comparison, we used three generative classifiers---the GMM, LDA, and Gaussian Naive Bayes (GNB)---and four discriminative classifiers---the $\nu$-SVM \citep{Chang2001-gd}, MLP, linear logistic regression (LLR), and $k$-NN.
For the classifiers involving hyperparameters, their values were optimized through five-fold cross-validation on the training data.
The number of components in the GMM was tuned in the range of 1--5.
The radial basis function was selected as the kernel function of $\nu$-SVM.
The hyperparameters $\nu_{\mathrm{SVM}}$ and $\gamma$ in the $\nu$-SVM were tuned by a 10 $\times$ 10 grid search ($\nu_{\mathrm{SVM}}$ ranging from $\log_{10} 0.99$ to $\log_{10} 1.0^{-5}$, and $\gamma$ ranging from $\log_{10} 5.0$ to $\log_{10} 1.0^{-5}$ in even intervals in logarithmic space).
The MLP had a single hidden layer and was trained with a batch size of 256 and a learning rate of 0.001.
We used a weight decay of $1.0 \times 10^{-5}$.
The number of nodes in the hidden layer of the MLP was tuned in the range of \{$D, D+2, D+4, \ldots, D+20$\}.
The value of $k$ in the $k$-NN algorithm was selected in the range of 1--10.
In the proposed method, the prior distribution settings and $\nu^{\mathrm{pre}}$ were the same as in the previous experiment, and the initial number of components was $K_c = 10$.
All algorithms including the proposed method were programmed using Python.
For the comparative methods, we used the implementation of the Scikit-learn package~\citep{scikit-learn}.
The experiments were run on a computer with an Intel Xeon W-3245 (3.2 GHz) processor and 96.0 GB RAM.

We calculated the performance of each method through four metrics: classification accuracy, tuning time, training time, and prediction time per record.
In general, the choice of trials used for training affects the classification performance.
To evaluate the classification performance under fair conditions, $s$ trials out of all the $T$ trials were used as the training set and the remaining $T - s$ trials were used as the test set.
The classification accuracy was calculated and averaged over all combinations for each participant.
Because it is difficult to procure many training examples in real-world applications, $s$ was set to be the largest integer less than or equal to $T/3$.
To save computation time, 5\% of the total training points were randomly selected for training.
The number of trials used for training and the sample size of the training/test data in each dataset is shown in Table~\ref{table:datasets}.
Note that the data lengths of Datasets IV--VI differed from trial to trial; therefore, the training and test sample sizes varied accordingly. 
Tuning time is defined as the total time taken for hyperparameter optimization based on cross-validation, and the training time is defined as the time required for the training of each model to converge.
Prediction time is the time taken to classify each test record (i.e., each data point).

\subsubsection{Comparison with previous studies}
We conducted a comparative experiment with previous studies.
The benchmark dataset used in this experiment was the Ninapro database 1~\citep{Atzori2014-nf}, which is available on Ninaweb\footnote{http://ninapro.hevs.ch/node/131/}.
This dataset contains 53 different upper-limb motions ($C=53$; including the rest state) recorded from 27 intact participants (age: $28.0 \pm 3.4$ years).
EMG signals were recorded using 10 EMG sensors ($D = 10$) at a sampling rate of 100 Hz.
Each motion consisted of 10 repetitions.
The signal segment labeled by each class contained a transient phase from the rest state to the motion state.

In this experiment, preprocessing, segmentation, and feature extraction were performed on the raw waveforms of the dataset according to the classification strategy proposed by Englehart and Hudgins~\citep{Englehart2003-ff} to match the conditions of previous studies to the maximum extent possible.
First, raw data were processed using a first-order low-pass filter with a cut-off frequency of 1 Hz~\citep{Atzori2014-nf} and were then segmented using a sliding window of 400 ms.
Thereafter, the mean absolute value, one of the representative amplitude features, was calculated for each segment and used as an input to the proposed method.
To evaluate the accuracy, motion repetitions \{1, 3, 4, 5, 9\} were used for training, and the remaining repetitions were used for testing~\citep{Atzori2015-dh,Cene2019-mb}.
The settings of the proposed method were the same as those in the previous experiments, except that the prior parameter of the Dirichlet distribution was changed to $\alpha_0 = 0.01$.
The reason for this change was to restrain the aggressive pruning of the model components and to maintain the model complexity as high as possible because the number of classes was significantly large in comparison with the number of electrodes (i.e., input dimension) in this dataset, thereby leading to complex class boundaries.

The results of the proposed method were compared with those obtained in the previous studies: random forest (RF)~\citep{Atzori2014-nf}, SVM~\citep{Atzori2015-dh}, convolutional neural network (CNN)~\citep{Atzori2016-cq,Geng2016-ug}, long short-term memory (LSTM) combined with MLP~\citep{He2018-ai}, and extreme learning machine (ELM)~\citep{Cene2019-mb}.
In these studies, the classification was performed under almost the same settings as those in this study.
Moreover, we did not employ post-processing techniques, such as majority voting, because the purpose was to purely compare the performance of the classifiers.

\section{Results}
\subsection{Simulation}
\begin{figure}[!t]
	\centering
  \includegraphics[width=0.75\hsize]{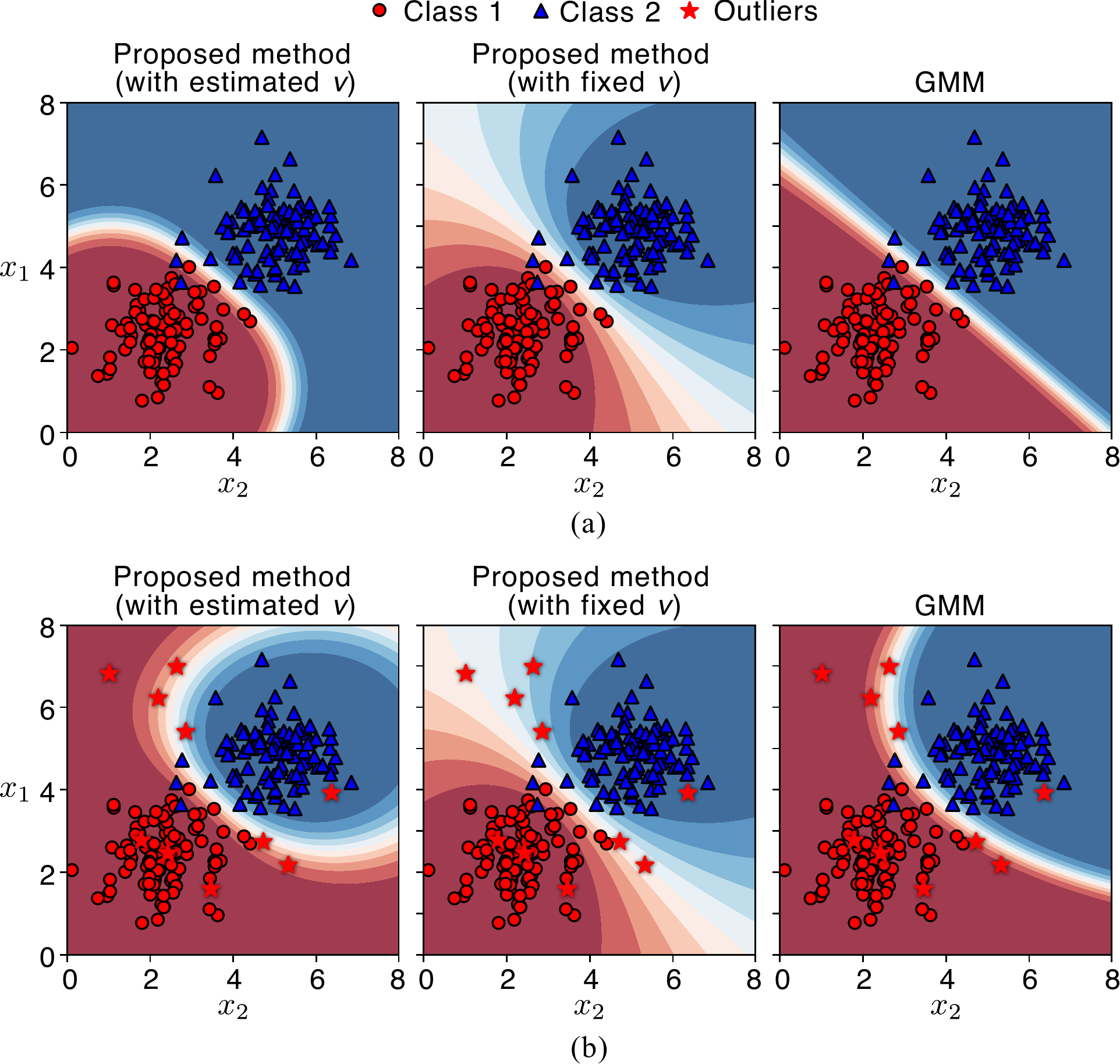}
	\caption{Scattergram of simulated data and color-coded posterior probability of each class. The left, middle, and right panels represent the results of the proposed method with $\nu$ estimated for each class, proposed method with $\nu$ fixed for all classes, and the Gaussian mixture model (GMM), respectively. (a) Simulated data without outliers. (b) Simulated data with outliers. The red and blue areas indicate posterior probabilities close to 1.0 for classes 1 and 2, respectively.}
	\label{fig:simulation}
\end{figure}

Fig.~\ref{fig:simulation} shows the color-coded decision boundary for each class given by the proposed method with $\nu$ estimated for each class, proposed method with $\nu$ fixed for all classes, and GMM.
%
%
The red and blue plots indicate labeled examples, and the red stars represent the added outliers for class 1.
In the proposed method (with estimated $\nu$) and the GMM, the classification boundaries changed significantly due to the outliers.

\subsection{EMG Analysis Experiments}

\begin{figure}[!t]
	\centering
  \includegraphics[width=0.7\hsize]{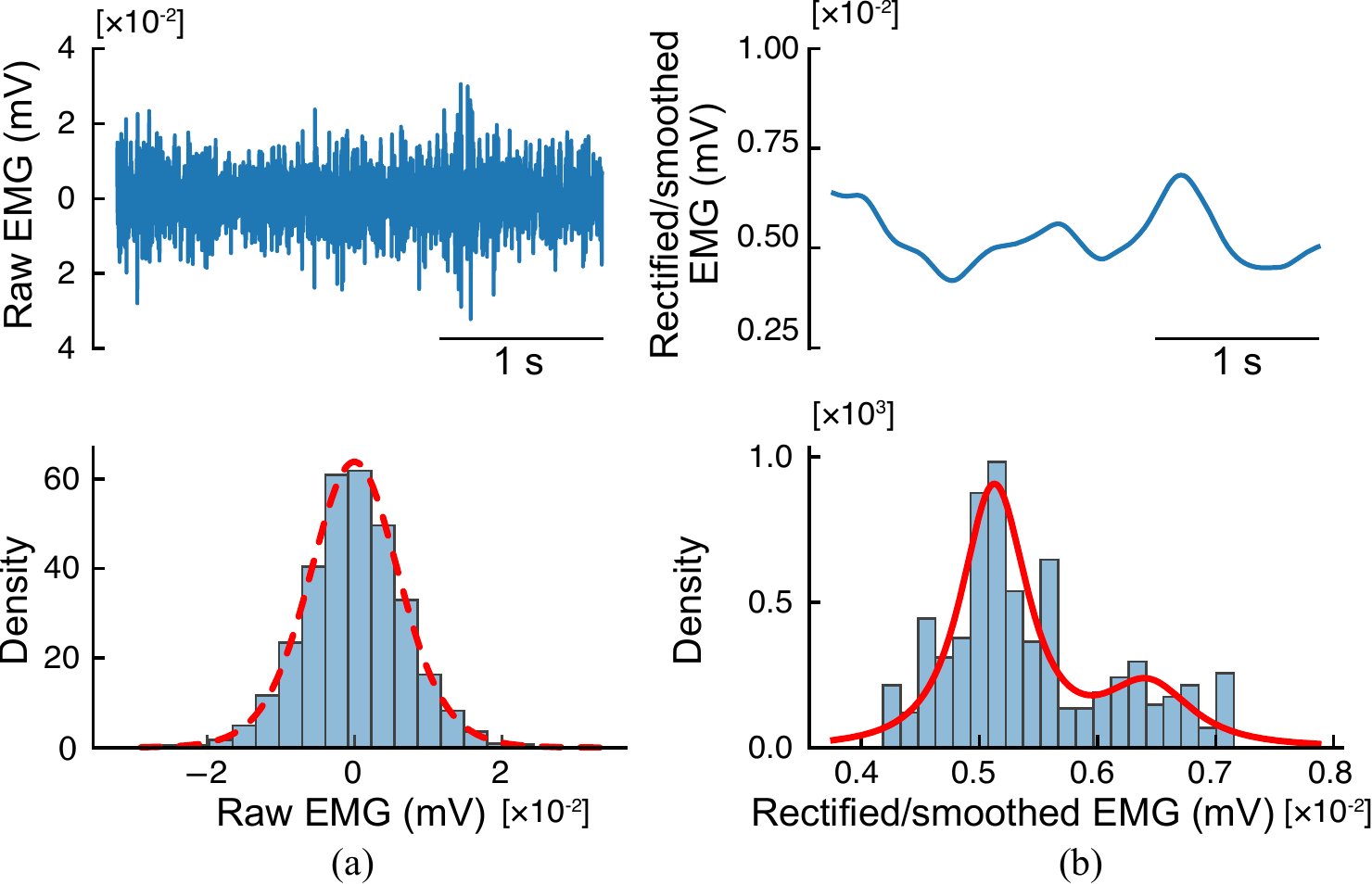}
	\caption{Raw and processed EMG signals and corresponding density histograms.
	This example was obtained from the second electrode for Participant 1 during hand supination.
	(a) Raw EMG signals. The dashed red line represents the fitting marginal distribution based on the scale mixture model~\citep{Furui2019-ho}.
	(b) Rectified and smoothed EMG signals. The solid red line represents the fitting marginal distribution based on the proposed finite mixture of scale mixture models with $\nu$ fixed at 2.
	}
	\label{fig:exp_dist}
\end{figure}

\begin{figure}[!t]
	\centering
  \includegraphics[width=0.65\hsize]{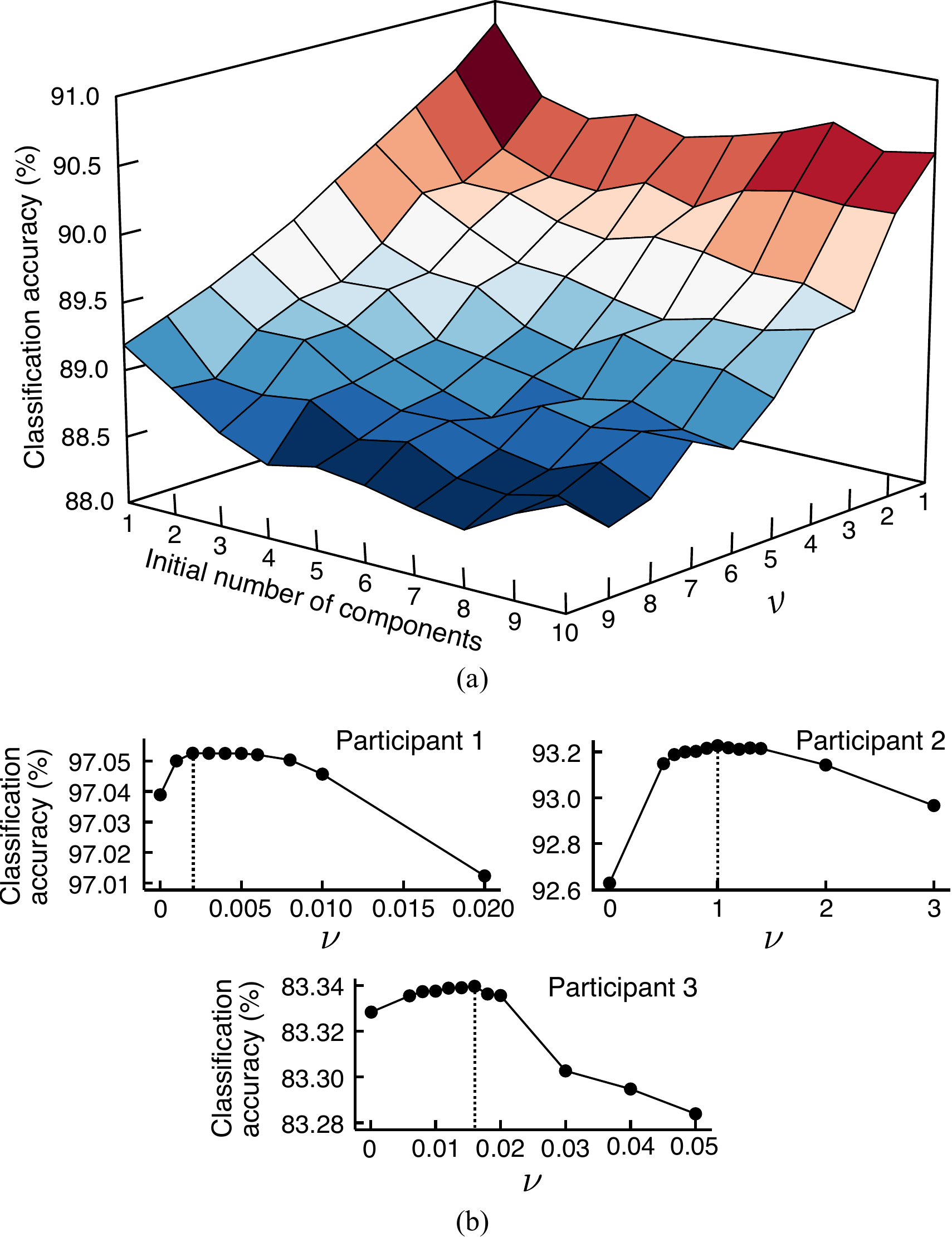}
	\caption{Evaluation results of the proposed method with various hyperparameters. (a) Average classification accuracy for each combination of the initial number of components $K_c$ and the parameter $\nu$. (b) Classification accuracy with changing $\nu$ when $K_c = 1$ for each participant.
	A detailed search was conducted around the maximum classification accuracy; the dotted lines indicate the points with the maximum accuracy.}
	\label{fig:grid}
\end{figure}

Fig.~\ref{fig:exp_dist} depicts examples of EMG signals and the corresponding density histograms from the second electrode for Participant 1 during hand supination.
It also shows the fitted distributions obtained using the original scale mixture model \citep{Furui2019-ho} for the raw EMG signals and the proposed Bayesian finite-mixture version for the rectified and smoothed EMG signals (Fig.~\ref{fig:exp_dist}(b)).
The initial number of components was set to 10, and two components remained after training.
%
%
Fig.~\ref{fig:grid}(a) presents the average classification accuracy of the proposed method over participants with various initial numbers of components $K_c$ and parameters $\nu$.
%
%
The average classification accuracy reaches a maximum value of $90.75$\% when $K_c = 1$ and $\nu = 1$.
The optimal value of $\nu$ that maximizes the classification accuracy differs depending on the participant (Fig.~\ref{fig:grid}(b)), with finer variations of $\nu$ at $K_c = 1$.
The participants exhibited the highest classification accuracy at $\nu = 0.002$ (Participant 1), $\nu = 1.0$ (Participant 2), and $\nu = 0.016$ (Participant 3).

\begin{figure}[!t]
	\centering
  \includegraphics[width=0.5\hsize]{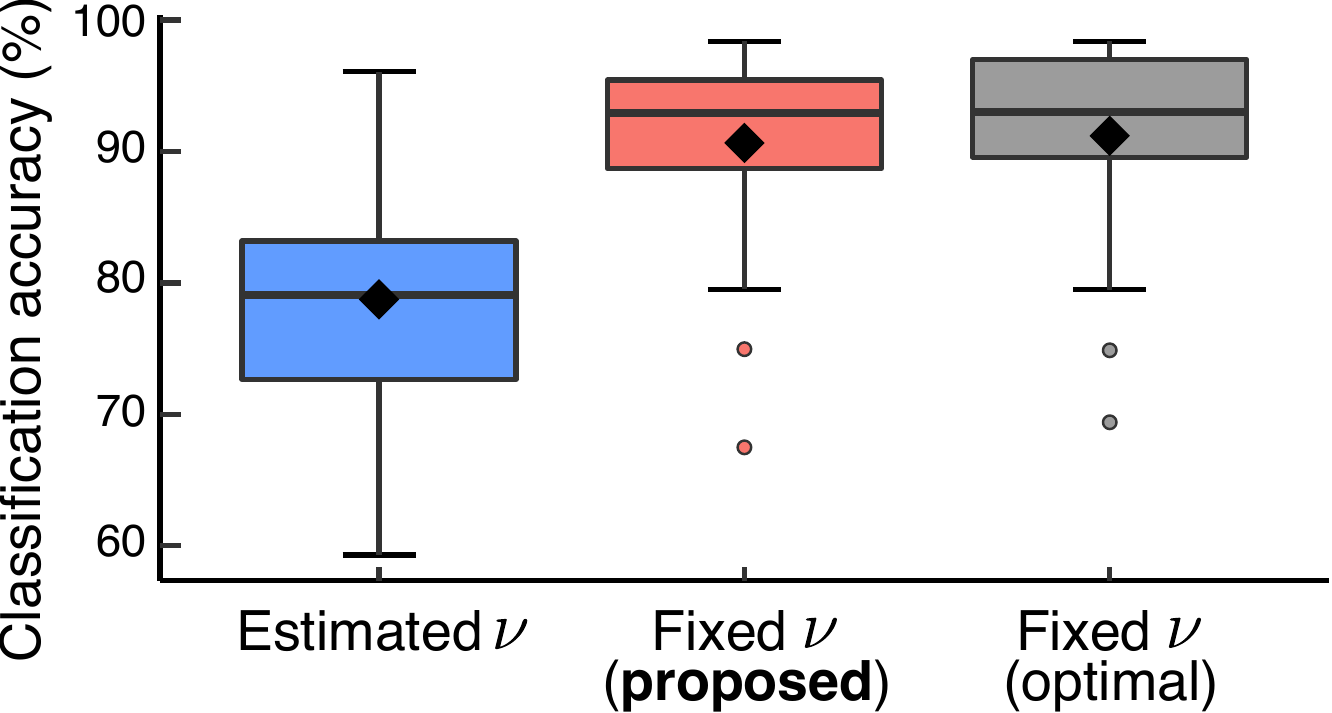}
	\caption{Classification results of the proposed method with different settings.
	The blue, red, and gray box plots represent the results with $\nu$ estimated for each class, $\nu$ fixed using the proposed maximum mutual information-based determination method, and $\nu$ fixed to its optimal value that maximizes the accuracy, respectively.
	The rhombuses in the box plots denote the average accuracies.}
	\label{fig:compare_grid}
	\vspace{-3mm}
\end{figure}

Fig.~\ref{fig:compare_grid} shows the classification accuracy for the proposed method with $\nu$ estimated for each class, $\nu$ fixed using the proposed maximum mutual information-based determination method, and $\nu$ fixed to the optimal value that maximizes the classification accuracy.
Owing to the small number of participants, statistical tests were not conducted.
%

%

\begin{figure}[!t]
	\centering
  \includegraphics[width=0.57\hsize]{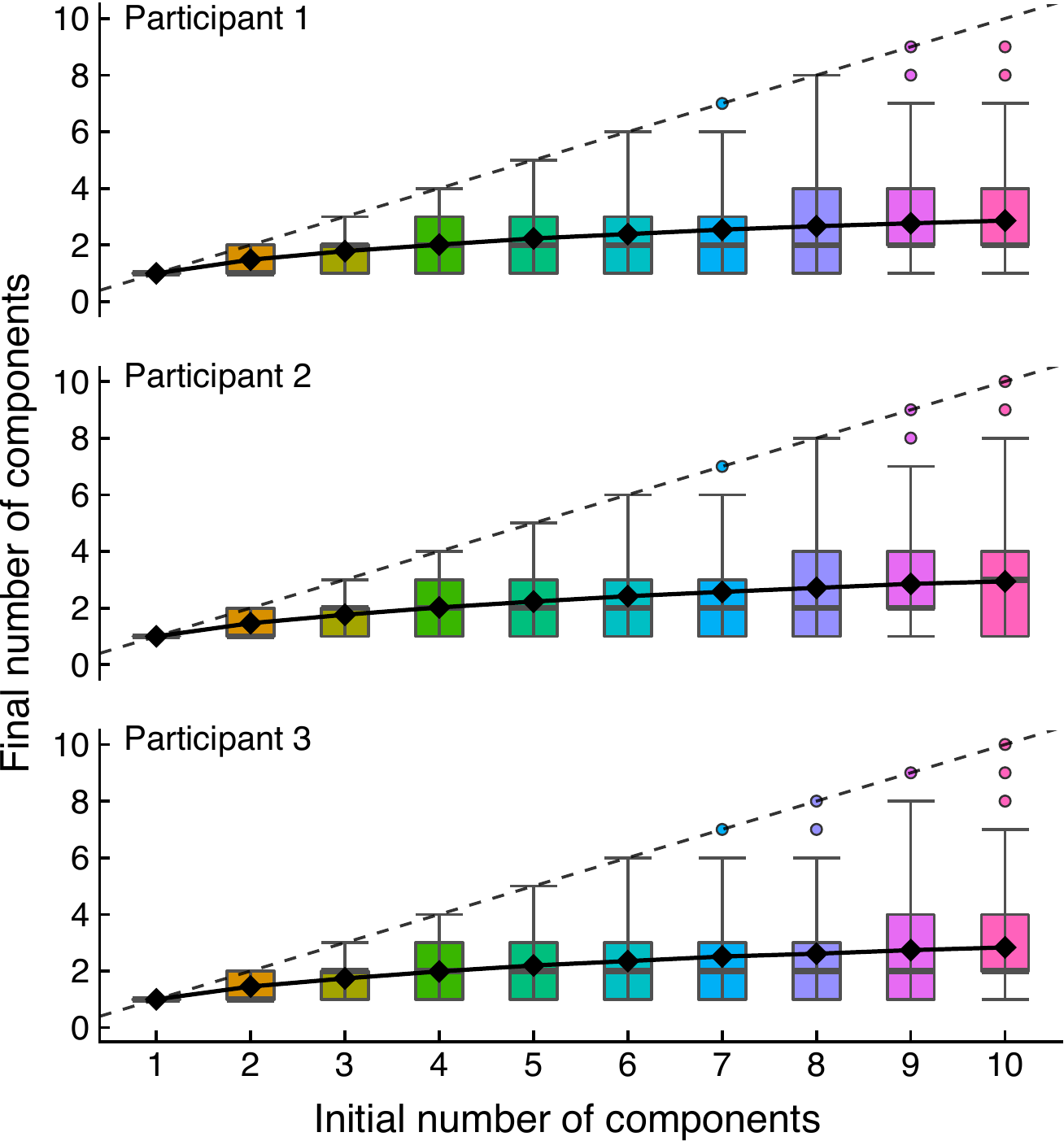}
	\caption{Number of components after training.
	The rhombuses in the box plots denote the average numbers of final components.}
	\label{fig:component_change}
\end{figure}

Fig.~\ref{fig:component_change} depicts the relationship between the number of initial and final components remaining after training. 
The dotted diagonal lines represent the points at which the numbers are equal.
For all participants, the average number of final components is less than 3, regardless of the initial number.
%
%

\subsection{EMG Classification Experiments}
Table~\ref{table:results} summarizes the results of EMG classification for each method.
The metrics are presented as \textit{average value $\pm$ standard deviation} for all participants.
The table also shows the significant differences in accuracy between the proposed method and comparative methods, as determined by performing a paired $t$-test with Holm adjustment.
The probability of superiority (PS)~\citep{Grissom1994-sp} is also given for accuracy to demonstrate the superiority of the classification on a per-participant basis.
The PS is an effect size that represents the probability of the experimental group showing a larger value than the control group (baseline), and can be easily calculated as $PS = N_{\mathrm{sup}} / N_{\mathrm{all}}$, where $N_{\mathrm{all}}$ is the number of all participants and $N_{\mathrm{sup}}$ is the number of times that the accuracy of the proposed method exceeds that of the comparative method in classification for each participant.
For example, $PS = 1.0$ indicates that the classification accuracy of the proposed method is superior to that of the comparative method for all participants; $PS < 0.5$ implies that the comparative method is more accurate than the proposed method for the majority of the participants.
In general, $PS > 0.56$, $PS > 0.64$, and $PS > 0.71$ are regarded as small, middle, and large effects, respectively~\citep{Grissom1994-sp}.

\begin{figure*}[t]
	\centering
  \includegraphics[width=1.0\hsize]{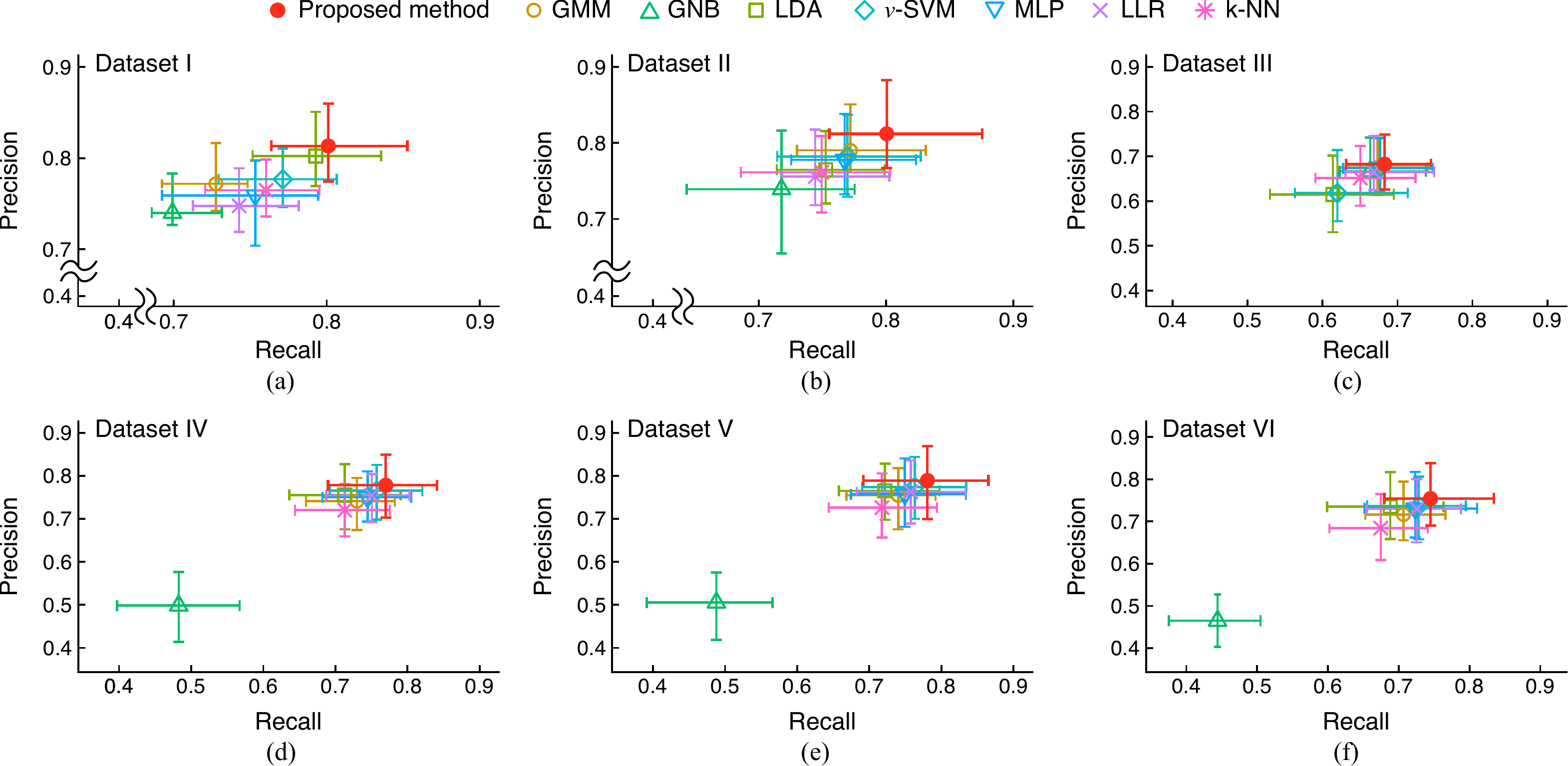}
	\caption{Precision versus recall for each classifier. 
	(a) Dataset I, (b) Dataset II, (c) Dataset III, (d) Dataset IV, (e) Dataset V, and (d) Dataset VI. 
	The scores of the proposed method are plotted as red circles.
	The error bars represent the 25th and 75th percentiles for all participants.}
	\label{fig:spec_recall}
\end{figure*}

Fig.~\ref{fig:spec_recall} shows the precision versus recall for each classifier and each dataset.
The precision and recall were calculated for each class separately and averaged.
The error bars represent the 25th and 75th percentiles for all participants.
It should be noted that the axes of the results for Datasets I and II are partially omitted due to the difference in the distribution of the values among the datasets.

\begin{table*}[tbp]
	\centering
	\caption{Results on EMG dataset classification}
	\begin{adjustbox}{totalheight=\textheight-2\baselineskip}
	\begin{threeparttable}
	\begin{tabular}{@{}clllll@{}}
	\toprule 
	\multicolumn{1}{l}{Dataset} & Method 	& Accuracy (\%)									&  Tuning time (s) 	&  Training time (s) 	&  \begin{tabular}[c]{@{}l@{}}Prediction time\\ per record (\si{\micro \second}) \end{tabular}  \\
	\midrule
	\multirow{9}{*}{I}			
								& GMM				& 72.75\,$\pm$\,9.37$^{*}$ {\footnotesize (PS $=$ 1.00)}		& 92.095\,$\pm$\,4.505		& 0.664\,$\pm$\,0.935 		& 1.827\,$\pm$\,0.902			\\
								& LDA				& 79.28\,$\pm$\,9.03$^{\ }$ {\footnotesize (PS $=$ 0.50)}	& 0							& 0.021\,$\pm$\,0.001		& 0.257\,$\pm$\,0.003		\\
								& GNB				& 69.91\,$\pm$\,9.62$^*$ {\footnotesize (PS $=$ 1.00)}		& 0							& 0.006\,$\pm$\,0.001		& 1.007\,$\pm$\,0.024			\\
								& $\nu$-SVM			& 77.13\,$\pm$\,9.50$^{\ }$ {\footnotesize (PS $=$ 0.88)}	& 4440.784\,$\pm$\,77.955	& 1.943\,$\pm$\,2.087		& 92.553\,$\pm$\,86.989		\\
								& MLP				& 75.33\,$\pm$\,8.75$^*$ {\footnotesize (PS $=$ 0.88)}		& 249.281\,$\pm$\,72.378	& 43.848\,$\pm$\,13.708		& 0.367\,$\pm$\,0.069		\\
								& LLR				& 74.27\,$\pm$\,9.61$^*$ {\footnotesize (PS $=$ 1.00)}		& 0							& 3.526\,$\pm$\,1.641		& 0.089\,$\pm$\,0.021			\\
								& $k$-NN			& 76.03\,$\pm$\,8.32$^*$ {\footnotesize (PS $=$ 0.88)}		& 147.092\,$\pm$\,4.711		& 0.011\,$\pm$\,0.001		& 30.136\,$\pm$\,1.471		\\
								\cmidrule(l){2-6}
								& Proposed & \textbf{80.10\,$\pm$\,8.42}		& 13.560\,$\pm$\,0.284	& 2.506\,$\pm$\,2.887				& 1.605\,$\pm$\,0.025			\\
	\midrule[0.04em]
	\multirow{9}{*}{II}			
								& GMM				& 77.20\,$\pm$\,9.20$^*$ {\footnotesize (PS $=$ 0.88)}		& 231.196\,$\pm$\,12.416		& 2.484\,$\pm$\,3.244				& 2.284\,$\pm$\,1.227			\\
								& LDA				& 75.27\,$\pm$\,8.65$^{\ }$ {\footnotesize (PS $=$ 0.88)}		& 0		& 0.049\,$\pm$\,0.002				& 0.268\,$\pm$\,0.004			\\
								& GNB				& 71.79\,$\pm$\,9.33$^*$ {\footnotesize (PS $=$ 0.88)}		& 0		& 0.012\,$\pm$\,0.002				& 1.070\,$\pm$\,0.004			\\
								&$\nu$-SVM			& 76.97\,$\pm$\,9.78$^{\ }$ {\footnotesize (PS $=$ 0.75)}		& 21444.220\,$\pm$\,452.622		& 12.673\,$\pm$\,14.975				& 276.724\,$\pm$\,269.778			\\
								& MLP				& 76.74\,$\pm$\,9.68$^{\ }$ {\footnotesize (PS $=$ 1.00)}		& 834.336\,$\pm$\,268.034		& 119.724\,$\pm$\,56.723				& 0.467\,$\pm$\,0.128			\\
								& LLR				& 74.44\,$\pm$\,10.43$^*$ {\footnotesize (PS $=$ 1.00)}		& 0		& 104.765\,$\pm$\,203.655				& 0.089\,$\pm$\,0.007			\\
								& $k$-NN			& 74.95\,$\pm$\,9.89$^{\ }$ {\footnotesize (PS $=$ 0.88)}		& 402.692\,$\pm$\,27.424	& 0.0325\,$\pm$\,0.006				& 35.384\,$\pm$\,3.224			\\
								\cmidrule(l){2-6}
								& Proposed 	& \textbf{80.05\,$\pm$\,9.46}		& 31.003\,$\pm$\,1.344	& 4.580\,$\pm$\,4.701			& 1.761\,$\pm$\,0.038			\\
	\midrule[0.04em]
	\multirow{9}{*}{III}		
								& GMM				& 68.55\,$\pm$\,8.34$^{\ }$ {\footnotesize (PS $=$ 0.75)}	& 30.454\,$\pm$\,3.858		& 0.436\,$\pm$\,0.536				& 0.832\,$\pm$\,0.454			\\
								& LDA				& 60.59\,$\pm$\,8.76$^{*}$ {\footnotesize (PS $=$ 1.00)}		& 0		& 0.005\,$\pm$\,0.001				& 0.201\,$\pm$\,0.021			\\
								& GNB				& 67.50\,$\pm$\,8.19$^{\ }$ {\footnotesize (PS $=$ 0.88)}		& 0		& 0.005\,$\pm$\,0.002				& 0.391\,$\pm$\,0.033			\\
								& $\nu$-SVM			& 62.58\,$\pm$\,11.52$^{*}$ {\footnotesize (PS $=$ 1.00)}		& 401.320\,$\pm$\,11.049		& 1.072\,$\pm$\,0.349				& 93.696\,$\pm$\,28.897			\\
								& MLP				& 68.74\,$\pm$\,8.84$^{\ }$ {\footnotesize (PS $=$ 0.75)}		& 128.976\,$\pm$\,16.450		& 21.094\,$\pm$\,6.072				& 0.285\,$\pm$\,0.047			\\
								& LLR				& 68.03\,$\pm$\,9.01$^{\ }$ {\footnotesize (PS $=$ 0.75)}	& 0		& 1.993\,$\pm$\,0.965				& 0.059\,$\pm$\,0.008			\\
								& $k$-NN			& 66.01\,$\pm$\,9.05$^*$ {\footnotesize (PS $=$ 1.00)}			& 74.261\,$\pm$\,1.167		& 0.002\,$\pm$\,0.000				& 20.288\,$\pm$\,0.456			\\
								\cmidrule(l){2-6}
								& Proposed 	& \textbf{69.28\,$\pm$\,8.71}		& 3.544\,$\pm$\,0.400	& 0.290\,$\pm$\,0.136				& 0.555\,$\pm$\,0.065			\\
	\midrule[0.04em]
	\multirow{9}{*}{IV}			
								& GMM				& 73.50\,$\pm$\,11.12$^*$ {\footnotesize (PS $=$ 0.93)}		& 15.434\,$\pm$\,9.168		& 0.096\,$\pm$\,0.078				& 0.632\,$\pm$\,0.270			\\
								& LDA				& 71.77\,$\pm$\,11.98$^{*}$ {\footnotesize (PS $=$ 0.93)}		& 0		& 0.003\,$\pm$\,0.001				& 0.173\,$\pm$\,0.021			\\
								& GNB				& 48.56\,$\pm$\,12.28$^*$ {\footnotesize (PS $=$ 1.00)}		& 0		& 0.001\,$\pm$\,0.000				& 0.384\,$\pm$\,0.035			\\
								& $\nu$-SVM			& 76.11\,$\pm$\,10.20$^{*}$ {\footnotesize (PS $=$ 0.70)}		& 48.476\,$\pm$\,49.219		& 0.095\,$\pm$\,0.124				& 17.646\,$\pm$\,14.905			\\
								& MLP				& 74.77\,$\pm$\,10.42$^*$ {\footnotesize (PS $=$ 0.72)}		& 102.729\,$\pm$\,45.663		& 14.827\,$\pm$\,5.383				& 0.290\,$\pm$\,0.064			\\
								& LLR				& 75.45\,$\pm$\,10.07$^*$ {\footnotesize (PS $=$ 0.70)}		& 0		& 1.303\,$\pm$\,1.991				& 0.047\,$\pm$\,0.006			\\
								& $k$-NN			& 71.64\,$\pm$\,10.04$^*$ {\footnotesize (PS $=$ 0.95)}		& 30.819\,$\pm$\,11.230		& 0.001\,$\pm$\,0.001				& 22.629\,$\pm$\,1.246			\\
								\cmidrule(l){2-6}
								& Proposed 	& \textbf{77.45\,$\pm$\,10.77}		& 1.672\,$\pm$\,0.512	& 0.702\,$\pm$\,0.502				& 0.567\,$\pm$\,0.059			\\
	\midrule[0.04em]
	\multirow{9}{*}{V}			
								& GMM				& 74.70\,$\pm$\,11.49$^{*}$ {\footnotesize (PS $=$ 0.93)}	& 15.199\,$\pm$\,7.787		& 0.093\,$\pm$\,0.080		& 0.498\,$\pm$\,0.179			\\
								& LDA				& 72.83\,$\pm$\,12.14$^{*}$ {\footnotesize (PS $=$ 0.91)}	& 0		& 0.003\,$\pm$\,0.001		& 0.149\,$\pm$\,0.025			\\
								& GNB				& 49.23\,$\pm$\,12.68$^{*}$ {\footnotesize (PS $=$ 1.00)}	& 0		& 0.001\,$\pm$\,0.000		& 0.305\,$\pm$\,0.034			\\
								& $\nu$-SVM			& 76.85\,$\pm$\,10.94$^{*}$ {\footnotesize (PS $=$ 0.82)}	& 49.141\,$\pm$\,49.197		& 0.091\,$\pm$\,0.117		& 17.060\,$\pm$\,14.225			\\
								& MLP				& 75.37\,$\pm$\,11.21$^*$ {\footnotesize (PS $=$ 0.89)}		& 104.438\,$\pm$\,45.089	& 14.786\,$\pm$\,5.309		& 0.255\,$\pm$\,0.038			\\
								& LLR				& 76.24\,$\pm$\,10.99$^*$ {\footnotesize (PS $=$ 0.84)}		& 0		& 1.166\,$\pm$\,1.653		& 0.032\,$\pm$\,0.005			\\
								& $k$-NN			& 72.29\,$\pm$\,11.55$^*$ {\footnotesize (PS $=$ 0.93)}		& 29.941\,$\pm$\,10.493		& 0.001\,$\pm$\,0.001		& 22.087\,$\pm$\,1.190			\\
								\cmidrule(l){2-6}
								& Proposed 	& \textbf{78.71\,$\pm$\,11.14}		& 1.743\,$\pm$\,0.590		& 0.696\,$\pm$\,0.493		& 0.483\,$\pm$\,0.077			\\	
	\midrule[0.04em]
	\multirow{9}{*}{VI}			
								& GMM				& 71.27\,$\pm$\,11.87$^*$ {\footnotesize (PS $=$ 0.95)}		& 14.631\,$\pm$\,6.921	& 0.098\,$\pm$\,0.060		& 0.500\,$\pm$\,0.195			\\
								& LDA				& 69.49\,$\pm$\,12.43$^{*}$ {\footnotesize (PS $=$ 0.97)}	& 0		& 0.003\,$\pm$\,0.001		& 0.151\,$\pm$\,0.023			\\
								& GNB				& 44.90\,$\pm$\,11.87$^*$ {\footnotesize (PS $=$ 1.00)}		& 0		& 0.001\,$\pm$\,0.000		& 0.263\,$\pm$\,0.027			\\
								& $\nu$-SVM			& 73.37\,$\pm$\,11.81$^{*}$ {\footnotesize (PS $=$ 0.75)}	& 55.918\,$\pm$\,50.964		& 0.110\,$\pm$\,0.133	& 18.908\,$\pm$\,16.031			\\
								& MLP				& 72.74\,$\pm$\,11.93$^*$ {\footnotesize (PS $=$ 0.72)}		& 108.698\,$\pm$\,47.470	& 15.083\,$\pm$\,5.430	& 0.263\,$\pm$\,0.039			\\
								& LLR				& 73.02\,$\pm$\,10.76$^*$ {\footnotesize (PS $=$ 0.72)}		& 0		& 1.442\,$\pm$\,2.138		& 0.033\,$\pm$\,0.005			\\
								& $k$-NN			& 67.99\,$\pm$\,12.16$^*$ {\footnotesize (PS $=$ 1.00)}		& 30.265\,$\pm$\,10.615		& 0.001\,$\pm$\,0.001	& 22.298\,$\pm$\,1.406			\\
								\cmidrule(l){2-6}
								& Proposed 	& \textbf{75.15\,$\pm$\,12.27}		& 1.748\,$\pm$\,0.596		& 0.715\,$\pm$\,0.488		& 0.431\,$\pm$\,0.072			\\						
	\bottomrule 
	\end{tabular}
	\begin{tablenotes}[para,flushleft]
		$^*$: significant difference from the proposed method as indicated by a paired $t$-test with Holm adjustment (\textit{p} $<$ 0.05). PS: probability of superiority of the proposed method. The highest accuracy for each dataset is indicated in \textbf{bold}.
		\end{tablenotes}
\end{threeparttable}
\end{adjustbox}
\label{table:results}
\end{table*}

Table~\ref{table:results_compare} presents the experimental results reported in previous studies as well as those achieved by the proposed method on the Ninapro database 1.
The results were averaged for all test repetitions; some previous studies lacked standard deviations.
The number of motion repetitions used for training differed for each study.
The proposed method yielded an accuracy of greater than 78\%, outperforming the methods proposed in previous studies.

\begin{table*}[t]
	\centering
	\caption{Results and comparison with related studies on Ninapro database 1}
	\begin{adjustbox}{max width=\textwidth-1.5cm}
	\begin{threeparttable}
	\begin{tabular}{llll}
	\toprule 
	\multicolumn{1}{l}{Method}      & Average accuracy (\%)   & \# Repetitions for training \\ 
	\midrule
	RF~\citep{Atzori2014-nf} 	 & 75.32 $\pm$ 5.69	& 7	\\
	SVM~\citep{Atzori2015-dh} 	 & 76.00	& 5 \\
	CNN~\citep{Atzori2016-cq} 	 & 66.59 $\pm$ 6.40	& 7	\\
	CNN~\citep{Geng2016-ug} 	 & 76.10	& 7 \\
	LSTM + MLP~\citep{He2018-ai} & 75.45 $\pm$ 8.97 & 5 \\
	ELM~\citep{Cene2019-mb} 	 & 75.03	& 5 \\
	\midrule[0.04em]
	Proposed	& \textbf{78.43 $\pm$ 4.51}	& 5 \\
	\bottomrule 
	\end{tabular}
	\begin{tablenotes}[para,flushleft]
		The highest accuracy is indicated in \textbf{bold}. In some studies, standard deviations were not provided.
	\end{tablenotes}
\end{threeparttable}
\end{adjustbox}
\label{table:results_compare}
\end{table*}


\section{Discussion}

This paper proposes an EMG pattern recognition method based on a stochastic generative model of EMG signals.
In the proposed method, the scale of the EMG signal distribution is considered as a latent random variable, thereby enabling classification by taking into account the uncertainty.
The proposed method involves two hyperparameters: the number of components $K_c$ and the degrees of freedom $\nu$.
To determine these parameters automatically during the training process, we introduced Bayesian inference and a mutual information-based determination method.
In the experiments, synthetic data and measured EMG signals were analyzed to investigate the characteristics of the proposed method.
The generalization performance in EMG classification of the proposed method was then evaluated using public EMG datasets and compared with that of the baseline methods.

In the simulation experiment, we investigated the effect of the presence/absence of outliers on the classification characteristics of the proposed method (Fig.~\ref{fig:simulation}). 
In the case of the proposed method with $\nu$ estimated for each class, the decision boundary was different between the cases with and without outliers, and the region of class 1 involving outliers was expanded.
In the proposed method, $\nu$ controls the tolerance of the dispersion in the EMG variance (i.e., the tailedness of the EMG distribution), and the distribution tail becomes heavier as $\nu$ decreases.
Therefore, $\nu$ is an extremely sensitive parameter, and if $\nu$ is estimated for each class based on a maximum-likelihood perspective, only certain classes with accidental variation due to outliers and other factors can have heavy tails, resulting in the expansion of the decision region.
For the GMM, the decision region of class 1 is also expanded after adding outliers, as the GMM cannot consider the variation in the variance and its covariance matrix is enlarged due to outliers.
In contrast, the proposed method with $\nu$ fixed for all classes drew decision boundaries linearly between the classes regardless of the presence or absence of outliers.
This is because, by setting a common $\nu$ for all classes, the tail weights of the distributions become equal, thereby preventing the over-expansion of the decision region for a certain class alone.
Therefore, the proposed method can perform robust training against accidental outliers by sharing $\nu$ among classes while considering the variation in the variance.
In practical applications, the proposed method with a fixed value of $\nu$ is expected to demonstrate better generalization performance as the variation in the variances of each class is likely to be different between training and prediction.

EMG analysis experiments were conducted to evaluate the applicability of the proposed method to actual EMG data.
In Fig.~\ref{fig:exp_dist}(a), the raw EMG signal follows a symmetrical distribution with a mean of zero, which is well fitted by the original scale mixture model.
In contrast, the distribution of EMG features after rectification and smoothing has a more complex shape involving skewness and multimodality (Fig.~\ref{fig:exp_dist}(b)).
The finite mixture of scale mixture models, included in the proposed method, can be flexibly fitted to such a complex distribution geometry.
These results suggest that the proposed method has sufficient representation capability to handle the processed EMG signals.

The classification performance of the proposed method was also evaluated while varying the hyperparameters, $\nu$ and the number of components.
The classification accuracy tended to increase with decreasing $\nu$ (Fig.~\ref{fig:grid}(a)), indicating that setting a small value of $\nu$ to allow more outliers provides better accuracy for unknown data.
A more detailed search for a value of $\nu$ that maximizes the classification accuracy revealed that the maximum accuracy points were distributed in the region $\nu \leq 1$ and varied across participants (Fig.~\ref{fig:grid}(b)).
These results suggest that $\nu$ needs to be set adaptively for each person.
In Fig.~\ref{fig:compare_grid}, the classification accuracy when $\nu$ is automatically determined for each participant using the proposed mutual information-based determination method is almost the same as that when the optimal value of $\nu$ is searched and fixed by performing a grid search.
Note that the latter method with optimal $\nu$ is unrealistic, as it is impossible to identify parameters that maximize accuracy for unknown test data in real-world applications.
This result implies that a quasi-optimal value of $\nu$ with high generalization performance can be obtained from only the training data by introducing the proposed determination method.
The accuracy was reduced by nearly 10\% when $\nu$ was estimated for each class based on the maximum likelihood method.
As discussed in relation to the simulation experiments, fixing $\nu$ for all classes significantly improves the generalization performance.

Next, the classification accuracy tended to be the highest when the number of initial components was 1 (Fig.~\ref{fig:grid}(a)).
When the number of initial components was greater than 2, the accuracy remained almost unchanged.
This tendency occurred because, as the number of initial components increased, the unnecessary components were pruned out from the model during the training process and eventually converged to a smaller number of components.
Although the final number of components after training tended to vary depending on the initial number of components, the mean and median final numbers of components were always less than 3 (Fig.~\ref{fig:component_change}).
Following an EMG pattern recognition study using a maximum-likelihood-based GMM, it was reported that a larger number of components caused overfitting to noise or outliers in the training data and decreased the generalization ability~\citep{Chan2003-hz}. 
In the proposed method, overfitting to such non-essential elements is suppressed by the automatic relevant determination based on Bayesian inference; hence, there is no significant decrease in accuracy even if the initial number of components is set to a redundant value.
Furthermore, a similar previous study using a GMM revealed that, although the accuracy tended to be higher as the number of components decreased, the optimal number of components differed depending on the participant~\citep{Huang2005-xi}.
Therefore, the optimal number of components was 1 in the present study, however, this value may change depending on the experimental configuration and participants.
These findings indicate that the proposed method can adaptively determine the number of components according to the complexity of the training data.
In this EMG analysis experiment, for simplicity, the data without the transient states between rest and motion were used.
Some attempts have been made to exclude the transient data from the classification using muscle force information estimated from EMG signals~\citep{Fukuda2003-ls,Furui2019-bz}; therefore, the present discussion is expected to be valid in real-world applications by incorporating such techniques.

The EMG classification experiment using public datasets revealed that the proposed method yielded a higher classification accuracy on average compared to those of the baseline methods (Table~\ref{table:results}).
The PS was always above 0.5 and was greater than 0.75 in most cases, implying that the proposed method also performed better in the participant-wise classification.
Furthermore, the proposed method was superior to the other methods in terms of both recall and precision (Fig.~\ref{fig:spec_recall}).
The SVM, MLP, and $k$-NN techniques are powerful discriminative classifiers and have been applied in a wide range of fields, including EMG classification.
The proposed method, which is a generative classifier, yielded a higher accuracy than those of the above-mentioned discriminative methods because it is based on a stochastic EMG model; therefore, the mismatch between the true data distribution and model was smaller.
These results suggest that even general-purpose classifiers such as the SVM and MLP are not always suitable for the particular application of EMG classification.
Although the GMM, LDA, and GNB approaches are generative classifiers, similar to the proposed method, they are based on simple Gaussian distributions.
Recent studies have suggested that EMG signals follow distributions that are more heavily tailed than Gaussian distributions, which is believed to be due to the variation in EMG variance caused by muscle force changes~\citep{Furui2019-ho} and muscle fatigue~\citep{Furui2019-os}.
The Gaussian distribution-based classifiers cannot take such variations into account, resulting in relatively low generalization performance.
In contrast, unlike the above-mentioned comparative methods, the proposed method is constructed based on an EMG model that can consider the stochastic variation in the variance; hence, the proposed method may be more suitable for EMG classification.
The effectiveness of the proposed method was also demonstrated through the comparison with previous studies (Table~\ref{table:results_compare}).
The proposed method yielded a higher accuracy than those of the well-known deep neural network models, namely CNN and LSTM.

To apply the classifier to an actual real-time system (e.g., myoelectric prosthesis), its computational efficiency, particularly the prediction time for novel data, is also important.
The experimental results indicate that the time required for the proposed method to predict the class label of a single data point is significantly short, less than 2.0 \si{\micro \second} for all datasets (Table~\ref{table:results}).
The training time is also relatively short, which is well within the acceptable range.
Although the proposed method requires additional tuning time to determine $\hat{\nu}$, its tuning time is considerably shorter than those of the other methods.
These results imply that the proposed method has high computational efficiency and can be applied to real-time systems.

Although the structure of the proposed method is generative, only $\nu$, which controls the tolerance of variations, was trained in a discriminative manner.
Such partial discriminative training may also contribute to the high generalization ability of the proposed method.
However, in the proposed method, generative and discriminative training are performed independently in separate procedures; hence, there may still be room for optimization. 
We believe that hybrid generative/discriminative approaches are effective for solving this problem.
Several reports have provided discriminative interpretations of probabilistic models~\citep{Tsuji1999-qb,Fukuda2003-ls,Klautau2003-xt}, and other studies have been conducted with the objective of integrating generative and discriminative approaches~\citep{Bishop2007-pd,Roth2018-gv}.
More consistent and effective inference methods based on such concepts are necessary to further improve the classification accuracy.

In the EMG pattern classification problem, training and test data are usually obtained from different recording sessions.
As the internal state of a person changes every moment, this problem inherently involves a covariate shift~\citep{Vidovic2014-wi}, which is the shift between the distribution of the training and test data.
The classification performance may decrease when the test distribution changes significantly because of electrode shifts, muscle fatigue, and posture changes~\citep{Betthauser2017-ct,Yang2019-xn,Ding2019-mr}.
To deal with these problems, some studies have introduced data augmentation for training data~\citep{Furui2017-su,Tsinganos2020-id} and adaptive inference by sequential learning~\citep{Vidovic2014-wi,Ding2019-mr}.
These approaches often involve heuristic settings and may require trial-and-error parameter tuning depending on the data.
As the proposed method is constructed as a Bayesian generative model, the new posterior distributions corresponding to the new data can be sequentially calculated by treating the old posterior distributions as prior distributions.
In addition, artificial data can easily be generated by the trained models using ancestral sampling.
Therefore, the proposed method has the potential to enable adaptation to the time-varying characteristics of the distribution and data augmentation in a unified Bayesian framework.

\section{Conclusion}
This paper proposed an EMG pattern classification method based on a scale mixture model of EMG signals. 
The stochastic model included in the proposed method is an extension of the original EMG model to a multidimensional and finite mixture model.
This allows the proposed method to take into account the uncertainty superimposed on the EMG signal, unlike in conventional classifiers.
The proposed model is trained by variational Bayesian learning, enabling the automatic determination of the number of mixture components by pruning out the redundant components.
In addition, we introduced a maximum mutual information-based determination method for the hyperparameter $\nu$. 
This method can train the degree of robustness to unknown data in a discriminative manner.

Simulation and EMG analysis experiments were conducted to evaluate the characteristics of the proposed method.
The results revealed that although the average classification accuracy tended to be high for small, fixed values of $\nu$, the optimal value of $\nu$ varied depending on the participant.
We also found that the hyperparameters could be optimized on a per-participant basis through the Bayesian learning and mutual information maximization introduced in the proposed method.
The comparison using public EMG datasets demonstrated that the proposed method outperformed general generative/discriminative classifiers.
These results indicate that accurate EMG pattern recognition can be achieved using the proposed method by considering the uncertainty in the EMG signals.

One of the limitations of this study is that we only verified the classification performance in an offline environment.
However, it is known that the accuracy in online environments tends to be higher than that in offline environments~\citep{Huang2010-jm}.
We therefore believe that the proposed method is also effective in online environments.
Nevertheless, to evaluate the applicability of the proposed method in practical situations precisely, it is necessary to conduct classification experiments in an online environment or to implement it on an actual prosthetic hand.

\section*{Acknowledgment}
This work was partially supported by a Grant-in-Aid for JSPS Research Fellow 18J22370 and 20K14698.

\bibliographystyle{model5-names}\biboptions{authoryear}
\bibliography{ref.bib}


\end{document}